\documentclass[]{article}
\RequirePackage[OT1]{fontenc}
\RequirePackage{amsthm,amsmath}
\RequirePackage[numbers]{natbib}
\RequirePackage[colorlinks,citecolor=blue,urlcolor=blue]{hyperref}

\usepackage{lscape}
\usepackage{lineno,hyperref}
\usepackage{amsfonts}
\usepackage{graphicx,psfrag,epsf}
\usepackage{enumerate}
\usepackage{framed}
\usepackage{url} 
\usepackage{flexisym}
\usepackage{subfig}
\modulolinenumbers[5]

\usepackage{graphicx}
\usepackage{subfig}
\usepackage{amsmath}
 \usepackage{amsfonts}
\usepackage{epsfig}
\usepackage{authblk}

\newcommand{\be}{\begin{equation}}
\newcommand{\ee}{\end{equation}}
\newcommand{\bes}{\begin{equation*}}
\newcommand{\ees}{\end{equation*}}
\newcommand{\beqn}{\begin{eqnarray}}
\newcommand{\eeqn}{\end{eqnarray}}
\newcommand{\beqns}{\begin{eqnarray*}}
\newcommand{\eeqns}{\end{eqnarray*}}

\newtheorem{theorem}{Theorem}

\newcommand{\bc}{\mathbf{c}}

\newcommand{\bu}{\mathbf{u}}
\newcommand{\by}{\mathbf{y}}

\newcommand{\bita}{\boldsymbol{\beta}}
\newcommand{\bepsilon}{\boldsymbol{\varepsilon}}
\newcommand{\balpha}{\boldsymbol{\alpha}}
\newcommand{\btheta}{\boldsymbol{\theta}}

\newcommand{\bx}{\mathbf{x}}

\newcommand{\bPhi}{\mbox{\mathversion{bold}$\Phi$}}

\newcommand{\bPsi}{\mbox{\mathversion{bold}$\Psi$}}

\begin{document}

%
\title{Simultaneous nonparametric regression \\ in RADWT dictionaries}
%
\author[1]{Daniela De Canditiis}
\author[2]{Italia De Feis}
\affil[1]{Istituto per le Applicazioni del Calcolo ``M. Picone" - Rome}
\affil[2]{Istituto per le Applicazioni del Calcolo ``M. Picone" - Naples}

%
%
%
%
%

\maketitle
\begin{abstract}

A new technique for nonparametric regression of multichannel signals is presented. The technique is based on the use of the Rational-Dilation Wavelet Transform (RADWT), equipped with a tunable Q-factor able to provide sparse representations of functions with different oscillations persistence. In particular, two different frames are obtained by two RADWT with different Q-factors that give sparse representations of functions with low and high resonance. 
It is assumed that the signals are measured simultaneously on several independent channels and that they share the low resonance component and the spectral characteristics of the high resonance component. Then, a regression analysis is performed  by means of the grouped lasso penalty.
Furthermore, a result of asymptotic optimality of the estimator is presented  using reasonable assumptions and exploiting recent results on group-lasso like procedures.
Numerical experiments show the performance of the proposed method in different synthetic scenarios as well as in a real case example for the analysis and joint detection of sleep spindles and K-complex events for multiple electroencephalogram (EEG) signals.
\end{abstract}

{\bf Keywords} RADWT, grouped LASSO, multichannel 

2010 MSC:  62G08 \hspace{0.1cm} 62G20 \hspace{0.1cm}	 62H12

\nolinenumbers

\section {Introduction}
\label{sec:intro}
\setcounter{equation}{0}

This paper deals with the problem of simultaneously recovering  $K$ different signals independently or simultaneously recorded under the hypothesis that these signals share common characteristics. Indeed,  
when drawing $K$ independent or simultaneous experiments over the same (unknown) causal relation among variables, we expect that changing the experiment should not affect the causal relation but only some experiment specific characteristics. 
This situation is typical in the biological field, where scientists make experiments with more replicas because they assume a causal relationship between genes and response common to all replicas, while retaining a replicate-specific variability, see \cite{He2016}, \cite{Ruffalo2017}, \cite{Yuan2016}, \cite{Anestis2011}; but commons characteristics are also expected in the medical field to model special EEG data, where one waits the simultaneous signals derive from the electrodes located in the subject's scalp at specific areas, see \cite{Selesnick2011}, \cite{Barros2000}, \cite{Parekh2017}. See \cite{Bobin2009} for many other examples applied to different signal and image processing problems.

Such kind of problem is addressed in many different research areas: in the machine learning community it is well known as the multi-task learning problem \cite{Liu2008}, \cite{Lozano2012} \cite{Pontil2008}, in the signal and image processing community as the multi-channel recovering problem \cite{Rakotomamonjy2011}, in econometrics as the panel-data problem, in the approximation theory as the conjoint analysis as well as in the mathematical statistics community it is a special case of the multivariate regression problem. The enormous interest which is growing around this problem is due to its flexibility in modeling different situations and in the possibility of using fast algorithm to solve it. 

In this paper we propose to treat the problem of simultaneous nonparametric regression from a new perspective by combining results from signal processing and statistical 
high-dimensional data analysis. In signal processing it is now well understood that orthogonal basis decompositions are not appropriate for signal recovery, since they can often fail to 
represent a particular function of interest efficiently, \cite{Donoho2003}. As a result, overcomplete representations such as wavelets and windowed Fourier expansions became mainstays of modern statistics and signal processing. Such representations are formalized through the theory of frames.  Frames can be generated by the action of operators on a template function (mother wavelet or Gabor atom), or be unstructured and random (as in compressive sensing). Here we use results about RADWT \cite{Selesnick2011}, which is a modern and fast computational tool for analyzing a very general class of signals.
In statistical high-dimensional data analysis it is established that the grouped-Lasso technique \cite{Yuan2006} for the selection and estimation of grouped variables is very effective to identify the dictionary elements that guarantee efficient estimation of the unknown regression function. The advantage of this approach is twofold. First, from a theoretical point of view, it is possible to control the estimation error by the so called {\it oracle inequalities}, and the error rate becomes nearly parametric providing the function of interest can be represented via a linear combination of just few dictionary elements satisfying certain assumptions. Second, from a computational point of view, the group gradient descendent method permits a very fast implementation of the optimization algorithm to find the optimal path.

The remainder of the paper is organized as follows. Section \ref{sec:model} describes the data model we are considering with the working  hypothesis. Section \ref{sec:inference} presents and discusses the inference procedure within the paradigm of group-lasso procedures, enlightening the connections with other existing procedures. Section \ref{sec:convergence} provides convergence results, while Section \ref{sec:simul_real} shows numerical experiments.

\section{The data model} 
\label{sec:model}

Consider the problem of recovering $K+1$ deterministic vectors $\bc$, $\bu^{(1)}$, \ldots, $\bu^{(K)}$ $\in \mathbb{R}^{n \times 1}$ from the following data
\be  \label{eq:general_model}
\by^{(k)}= \bc +\bu^{(k)} +\bepsilon^{(k)}   \quad \quad  \quad k=1,...,K \quad \mbox{and} \quad \bepsilon^{(k)} \sim N(\mathbf{0}, \sigma^2 I)
\ee
where vector $\by^{(k)}$ represents $n$-equispaced observations of function $c(t)+u^{(k)}(t)$ over the equispaced grid design $ t_1 < t_2< \cdots < t_n$ for each channel $k=1,...,K$, i.e. $\by^{(k)} \in \mathbb{R}^{n \times 1}$. The grid can be thought to be sampled in time, in space, in radiation, in genome locations or in any other unit of measure according to the physical phenomena. The data model (\ref{eq:general_model}) represents the situation where the samples share a common effect, here represented by function $\bc(t)$ which eventually can be zero, plus a functional component $u^{(k)}(t)$ which can be different across samples while sharing some common characteristics to be specified later. We do not hypothesize  functions $c(t)$ and $u^{(k)}(t)$ belong to some functional Sobolev space $H^s_{p,q}[a,b]$ as it is usually done in functional nonparametric regression setting, instead we let these functions to be much more general and we restrict our attention to their finite-dimensional representation.  Since many physiological and physical signals are not only non-stationary but also  exhibit  a  mixture  of  oscillatory  and  non-oscillatory transient behaviors (for example, speech, stock-market, biomedical EEG,  etc) we suppose that each signal in each channel is the sum of a `high-resonance' and a `low-resonance' component.  By a high-resonance component, we mean a signal consisting of multiple simultaneous sustained oscillations, in  contrast,  by  a  low-resonance  component,  we mean  a  signal  consisting  of  non-oscillatory  transients of unspecified shape and duration.  We stress that the high and low resonance component of a signal can not be extracted from its high and low frequencies components in a time-scale decomposition, but they can be well represented by a high-Q factor RADWT and a low-Q factor RADWT respectively as very well explained in \cite{Selesnick2011}.  The RADWT is a normalized tight frame of $L_2 (R)$  defined as
$\left\{ (\frac{q}{p}) ^{k/2}\psi \left( (\frac{q}{p})^k t+\frac{sp}{q} l \right)  \right\}_{k,l \in Z}$ where $\psi$
is a wavelet function and ($p,q,s$) is a triplet of parameters which gives the time-scale characteristic of the frame. In particular the ratio  $ q/p > 1$ is closely related to the scale (or frequency) dilatation factor, the parameter $s$ is closely related to the time dilatation factor and $\frac{p}{s(q-p)}$ is the redundant factor. The Q-factor depends on these parameters although there is
not a an explicit formula, in particular setting the dilatation factor $q/p$ between 1 and 2 and $s>1$ gives a RADWT with high Q-factor, while setting $s=1$ we obtain a low Q-factor RADWT 
with time-scale characteristic similar to the dyadic wavelet transform. In particular, when $q=2$, $p=1$ and $s=1$ the frame reduces to the classical wavelet basis. Given a finite energy 
signal $\bx$ of length $n$ and $J \in \mathbb{N}$  levels of decomposition, the RADWT transform is obtained by a sequence of proper down-sampling operations and fast Fourier transforms; it ends up with $\lceil \frac{n p^J}{q^J} \rceil$ scaling coefficients (low-pass filtering) and $\lceil \frac{n p^j}{q^j s}\rceil $  wavelet
coefficients (high-pass filtering) at each level $j=1,..J$. See  \cite{Bayram2009} for details on fast analysis and synthesis schemes. In this paper we use these results of signal processing in order to formulate our working hypothesis. Let $\boldsymbol{\Psi} \in \ensuremath{{\mathbb R}}^{n \times d1}$ be the finite matrix representation of the low Q-factor analysis filter and let $\boldsymbol{\Phi} \in \ensuremath{{\mathbb R}}^{n \times d2}$ be the finite matrix representation of the high Q-factor analysis filter (the synthesis operators being just the transpose matrices), then our working hypothesis is the following:

\begin{itemize}

\item[(H1)] signal $\bc$  is sparse in $\boldsymbol{\Psi}$, i.e. setting $\balpha_{0}=\Psi^t \bc$ we have that  $\left|S_0^{\balpha}\right|=\left| \{j: \alpha_{0_j} \neq 0  \}\right|<< d1$;

\item[(H2)] signals $\bu^{(k)}$ have a {\bf jointly} sparse representation in $\boldsymbol{\Phi}$, i.e. setting $\bita^{(k)}_0=\boldsymbol{\Phi}^t \bu^{(k)}$ and $S_0^{(k),\bita}=\{ j : \beta^{(k)}_{0_j} \neq 0 \}$ we have that $ S_0^{(1),\bita}=\dots=S_0^{(K),\bita}$, with the common cardinality denoted by $\left|S_0^{\bita}\right|<< d2$.

\item[(H3)] the columns of matrices $\boldsymbol{\Psi}$ and $\boldsymbol{\Phi}$ are normalized to have norm 1.

\end{itemize}

Finally it is worth to observe that the role of  $\boldsymbol{\Psi}$ and $\boldsymbol{\Phi}$ in this model can be interchanged to accomplish cases where the common effect $\bc$ has a high Q-factor behaviour as opposed to the sample specific effect which has a low Q-factor behaviour.

\section{Inference} 
\label{sec:inference}
The linear model in (\ref{eq:general_model}) can be rewritten in terms of  RADWT coefficients as follows
\begin{eqnarray} \label{eq:model_channel}
\by^{(1)}&=& \boldsymbol{\Psi} \balpha + \boldsymbol{\Phi} \bita^{(1)}+\bepsilon^{(1)} \nonumber \\
\nonumber \\
\by^{(2)}&=& \boldsymbol{\Psi} \balpha + \boldsymbol{\Phi} \bita^{(2)}+\bepsilon^{(2)}  \\   
& \vdots & \nonumber \\
\by^{(K)}&=& \boldsymbol{\Psi} \balpha + \boldsymbol{\Phi} \bita^{(K)}+\bepsilon^{(K)} \nonumber  \\ \nonumber
\end{eqnarray}
which turns out to be a classical multiple regression model with a special common design matrix.
A first and somewhat naive approach would consist in treating separately each channel ignoring the underlying common structure; however this is obviously suboptimal. This is the reason why  such kind of problem is reformulated in terms of a unique regression problem in the following form:

\be   \label{eq:general_linear_model}
\begin{array}{c}
\left[\begin{array}{c}
\by^{(1)} \\
\by^{(2)} \\
\vdots \\
\by^{(K)} \\
\end{array}   \right]  
=
\left[ \begin{array}{ccccc}
\boldsymbol{\Psi} &\boldsymbol{\Phi}  & \boldsymbol{0} & \cdots & \boldsymbol{0} \\
\boldsymbol{\Psi} & \boldsymbol{0} & \boldsymbol{\Phi}  & \cdots & \boldsymbol{0} \\
\cdots & & & \cdots \\
\boldsymbol{\Psi}& \boldsymbol{0} &\boldsymbol{0} & \cdots & \boldsymbol{\Phi}  \\
\end{array}   ~
 \right] 
 \left[\begin{array}{c}
\balpha \\
\bita^{(1)} \\
\bita^{(2)} \\
\vdots \\
\bita^{(K)} \\
\end{array}   \right] 
+
\left[\begin{array}{c}
\bepsilon^{(1)} \\
\bepsilon^{(2)} \\
\vdots \\
\bepsilon^{(K)} \\
\end{array}   \right] =X  ~ \btheta +\bepsilon
\end{array}
\ee
with obvious correspondence between elements of the two expression. So, $\by$ is a column vector of $nK$ response variables, $\boldsymbol{X} $ a design matrix of dimension $nK \times d_1+ Kd_2$, $\btheta$ an unknown regression coefficients column vector of length $ d_1+ Kd_2$ consisting of a first sub vector $\balpha \in \mathbb{R}^{d_1 \times 1}$ and a second sub vector $\bita=\left[\left(\bita^{(1)}\right)^t, \ldots, \left(\bita^{(K)}\right)^t\right]^t \in \mathbb{R}^{K d_2 \times 1}$ and, finally, we let $\bepsilon$  be a $nK$-variate Gaussian random column vector with zero mean and  covariance matrix $\sigma^2 \boldsymbol{I}_{nK}$. 
Under the working hypothesis (H1) and (H2),  we expect the coefficients of the common part $\balpha$ to be sparse into the dictionary $\boldsymbol{\Psi}$, while on the remaining part of coefficient vector $\bita$ we exploit the joint sparsity assumption, i.e. for all $j=1,...,d_2$ we know that $\beta_j^{(k)}=0$, for all $k=1,..,K$ or $\beta_j^{(k)} \neq 0$ for all $k=1,...,K$. This provides the following non-overlapping group structure for the whole vector 
$\btheta=\left[\begin{array}{c}
\balpha \\
\bita\\
\end{array}   \right]$:

\be \label{eq:gruppi}
\{1,2...,d_1 + K d_2 \}=\{1\} \cup \dots \cup \{d_1 \} \cup G_1 \cup \cdots \cup G_{d_2}, 
\ee
with
\bes
 \quad  G_j=\{d_1+j, d_1+j+d_2, d_1+j+2 d_2 \ldots, d_1+j+(K-1)d_2 \}, j=1,\dots,d_2
\ees
group of size $K$.
Let   $G^{\star} =\frac{d_1+ K d_2}{d1+d2}$ denote the average group size and let us denote 
$$
\| \btheta  \|_{2,1}=\left \| \left[\begin{array}{c}
\balpha \\
\bita\\
\end{array}   \right] \right\|_{2,1} = \sqrt{\frac{1}{G^{\star}}} \sum_{j=1}^{d_1} |\alpha_j| +  \sqrt{\frac{K}{G^{\star}}} \sum_{j=1}^{d_2} \| \bita(G_j) \|_2
$$
the $l_1 / l_2$-norm, with $\bita(G_j)$ denoting the reduction of vector $\bita$ to the subset of index $G_j$, then we can consider the following group lasso problem   
\be  \label{eq:estimator_grouped}
\hat{\btheta}=\underset{\btheta \in \ensuremath{{\mathbb R}}^{(d1+K d_2)\times 1}}{argmin}  \left\{ \frac{1}{nK} || \by -\boldsymbol{X} \btheta ||_2^2 + \lambda \sqrt{G^{\star}} ||\btheta||_{2,1}    \right\}
\ee

Finally, we consider as our estimator the following reconstructions:
\begin{equation}
\hat{\boldsymbol{c}}=\boldsymbol{\Psi} \hat{\balpha}; \quad \hat{\boldsymbol{u}}^{(k)}=\boldsymbol{\Phi} \hat{\bita}^{(k)}, ~~~ k=1,\ldots,K \label{eq:estimator_final}
\end{equation}
where $\hat{\btheta}=\left[\begin{array}{c}
\hat{\balpha} \\
\hat{\bita}\\
\end{array}   \right] =\left[\hat{\balpha}^t,\left(\hat{\bita}^{(1)}\right)^t, \ldots, \left(\hat{\bita}^{(K)}\right)^t\right]^t $ is the solution of the optimization problem (\ref{eq:estimator_grouped}).

\subsection{Algorithm}
As already mentioned in the introduction one of the great advantages of the grouped Lasso penalization consists in the availability of efficient algorithms for its solution.

In particular, the most efficient algortihms in the  modern statistics literature are the Group Descendent Algorithm, presented in \cite{Breheny2009} and  \cite{Breheny2015} and implemented in the R package {\tt grpreg} available at https://cran.r-project.org/web/packages/grpreg/, and the Groupwise Majorization Descendent Algorithm presented in \cite{Yang2015} and implemented in the R package {\tt gglasso} available at https://cran.r-project.org/web/packages/gglasso/. 
Both algorithms work groupwise by using the separability of model (\ref{eq:estimator_grouped}), i.e. update each group of variables iteratively until convergence. The main difference between the two algorithms is the updating of each group of variables: in { \tt grpreg}  it occurs through the solution of a single-group lasso, i.e. with a multivariate soft-thresholding operator, under the assumption of ``othonormal group", while in {\tt gglasso} each group of variable is updated as the solution of a quadratic majorization problem. We stress that the ``orthonormal group" property refers to the condition $\boldsymbol{X}(G_j)^t \boldsymbol{X}(G_j)=I$, not that groups $\boldsymbol{X}(G_j)$ and $\boldsymbol{X}(G_k)$ are orthogonal each other. When this condition is not satisfied the {\tt grpreg} automatically orthonormalizes the design matrix, but this practice leads to a slight modification of the $l_1 / l_2$-norm contained in the penalty, as pointed out in \cite{Huang2012} and \cite{Simon2012}. 
This is not our case, because the design matrix defined in Eq. (\ref{eq:general_linear_model}) satisfies the ``orthonormal group" property and we can take complete advantage of the Group Descendent Algorithm in the {\tt grpreg} package to solve problem (\ref{eq:estimator_grouped}) exactly.

Let us reorganize the design matrix $\boldsymbol{X}$ defined in Eq. (\ref{eq:general_linear_model}) so that the group memberships are consecutive.
From the group structure defined in Eq. (\ref{eq:gruppi}) we have that the group membership vector $I_g$ contains only one element for $g=1,2,...,d_1$, and $K$ elements for $g=d_1+j$ with $j=1,\dots,d_2$. Hence, in the latter case the sub matrix $\boldsymbol{X}_{I_g}$, for $g=1,...,d_1$, is a one-column matrix defined as
$$
\boldsymbol{X}_{I_g} = 
\left[ \begin{array}{c}
\boldsymbol{\Psi}^{(g)} \\
\boldsymbol{\vdots} \\
\boldsymbol{\Psi}^{(g)}\\
\end{array}   ~
 \right] \in R^{nK\times 1},
$$
where $\boldsymbol{\Psi}^{(g)}$ is the $g$-th column of matrix $\boldsymbol{\Psi}$; 
while in the last case, for $g=d_1+j$ with $j=1,...,d_2$, the sub matrix $\boldsymbol{X}_{I_g}$ is a $K$-column matrix where each column is a shifted version of the $j$-th column of matrix 
$\boldsymbol{\Phi}$ as in the following scheme

$$
\boldsymbol{X}_{I_g} = 
\left[ \begin{array}{ccccc}
\boldsymbol{\Phi}^{(j)} & \boldsymbol{0} & \cdots & \boldsymbol{0} \\
 \boldsymbol{0} & \boldsymbol{\Phi}^{(j)} & \cdots & \boldsymbol{0} \\
\cdots & \cdots & \cdots & \cdots \\
\boldsymbol{0}& \boldsymbol{0} & \cdots & \boldsymbol{\Phi}^{(j)} \\
\end{array}   ~
 \right] \in R^{nK\times K}.
$$
Finally, it is easy to verify the ``orthonormal group" property, i.e. $\boldsymbol{X}_{I_g}^t \boldsymbol{X}_{I_g}=I$ for all $g=1,...,d_1+d_2$.

\subsection{Connections with literature}
As already stated in the introduction, multi-channel regression and equivalent problems have been investigated by diverse communities and a lot of literature is available on that. 

Problem  (\ref{eq:model_channel}) is a particular case of the so-called Simultaneous Sparse Approximation $(SSA)$ (\citep{Rakotomamonjy2011}, \citep{Jenatton2011}, \citep{Tropp2006_1}, \citep{Tropp2006_2}), defined as follows. Suppose that we have measured $K$ signals $\left\{\mathbf{s_i}\right\}_{i=1}^K$, where each signal is of the form $\mathbf{s}_i=\boldsymbol{\Omega c}_i+\bepsilon^{(i)}$, where $\left\{\mathbf{s}_i\right\} \in \mathbb{R}^{n \times 1}$, $\boldsymbol{\Omega}  \in \mathbb{R}^{n \times m}$ is a matrix of unit-norm elementary functions, $\boldsymbol{c}_i  \in \mathbb{R}^{m \times 1}$ is a weighting vector and $\bepsilon^{(i)}$ is a noise vector for each $i=1,\dots,K$. The overall measurements can be written as
\begin{eqnarray}
\mathbf{S}=\boldsymbol{\Omega C}+\bepsilon \label{SSA}
\end{eqnarray}
where $S=\left[\mathbf{s}_1,\dots,\mathbf{s}_K\right]$ is a signal matrix, $C=\left[\mathbf{c}_1,\dots,\mathbf{c}_K\right]$ a coefficient matrix and $\bepsilon$ a noise matrix. For the SSA problem, the goal is then to recover the matrix $\mathbf{C}$ given the signal matrix $\mathbf{S}$ and the dictionary $\boldsymbol{\Omega}$ under the hypothesis that all signals $\mathbf{s}_i$ share the same sparsity profile. This latter hypothesis can be translated into the request that the coefficient matrix $\mathbf{C}$ has a minimal number of non-zero rows, i.e. solving the following problem
\begin{eqnarray*}
\min_{\mathbf{C}}\frac{1}{2}\left\| \mathbf{S}-\boldsymbol{\Omega C}\right\|^2_F \quad s.t. \quad \left\| \mathbf{C}\right\|_{row-0} \leq T
\end{eqnarray*}
where
\begin{eqnarray*}
\left\| \mathbf{C}\right\|_{row-0}=\left| \{i \in \left[1,\dots,m\right]: c_{ij} \neq 0  ~ \mbox{for some j}\}\right|,
\end{eqnarray*}
$T$ is some parameter defined by the user to control the sparsity and $\left\| \cdot\right\|_F$ indicates the Frobenius norm.

This problem is not convex, but efficient greedy algorithms have been proposed  to get an approximate solution.  In particular, in \cite{Tropp2006_1}, the author proposes the Simultaneous Orthogonal Matching Pursuit (SOMP) algorithm, which selects, at each iteration, an element from the dictionary maximizing the sum of the absolute correlation between the dictionary elements and the signal residual. As shown in \cite{Rakotomamonjy2011}, this greedy algorithm is actually one of the most efficient to solve the problem. 

Another possibility to solve the minimization problem is to relax the constraint by replacing $\left\| \mathbf{\cdot}\right\|_{row-0}$ with a more tractable row-sparsity measure. A large class of relaxed version of $\left\| \mathbf{\cdot}\right\|_{row-0}$ consider the following constraint
\begin{eqnarray*}
J_{p,q}\left(\mathbf{C}\right)=\sum_i \left\|  \mathbf{c}_{i,\cdot} \right\|_q^p \quad \mathrm{with} \quad \left\|  \mathbf{c}_{i,\cdot} \right\|_q =\left(\sum_j \left| c_{i,j}\right|^q\right)^{1/q}
\end{eqnarray*}
where tipically $p \leq 1$ and $q \geq 1$.  

Such kind of relaxed problems can be solved in different ways and a deep survey and comparison analysis can be found in \citep{Tropp2006_2} and \cite{Rakotomamonjy2011}.

In particular, the case $p=1$ and $q=2$ can be efficiently solved by the Block Coordinate Descent (BCD) algorithm and has a strong connection with the group-lasso regression.  Indeed, our problem (\ref{eq:general_model}) falls in this relaxed version, considering 
\begin{eqnarray*}
\mathbf{S}=\boldsymbol{Y}=\left[\by^{(1)} \dots \by^{(K)}\right], \quad \boldsymbol{\Omega}=\left[ \bPsi, \bPhi \right] \quad \mathrm{and} \quad 
\mathbf{C}=\left[ \begin{array}{c}
\balpha^{(1)} \\ \bita^{(1)} \end{array}   \dots
 \begin{array}{c} \balpha^{(K)} \\ \bita^{(K)} \end{array} \right].
\end{eqnarray*}

Moreover, there is also a connection with structured variable selection and structural penalties in the vector formulation of Eq. (\ref{eq:general_linear_model}). Infact, the penalty we used in Eq. (\ref{eq:estimator_grouped}) is a particular case of Eq. (1), Section 2, described in \cite{Jenatton2011}, and this permits to use all the optimization algorithms based on the proximal methods. 

Finally, it is important to stress a fundamental difference with the proposed methodology, i.e. all reviewed methods don't take properly into account the constraint of a common low-component $\left(\balpha^{(1)}=...=\balpha^{(K)} \right)$, hence any multichannel reconstruction returns different low-resonance components for different channels, loosing in term of estimation error as it will be shown into the numerical section.

\section {Theoretical properties}
 \label{sec:convergence}
The following results are obtained adapting results of Chapter 8 in \cite{geer_book}.  

Let estimator $\left[\hat{\boldsymbol{c}}^t, \left(\hat{\boldsymbol{u}}^{(1)}\right)^t,\ldots,\left(\hat{\boldsymbol{u}}^{(K)}\right)^t\right]^t$ be given by Eq. (\ref{eq:estimator_final}); in order to derive a oracle inequality for its error, we introduce the following notations and assumptions.

\vspace{2mm}

\noindent
{\bf Notations:} \ for any subset of indices  $S \subseteq \mathcal{P} =\{1, \ldots,d_1 \} \cup \{d_1+1,\ldots,d_1+d_2\}$, we denote $S^{\balpha}=S \cap \{1,\ldots,d_1\}$ and $S^{\bita}=\{j: 1\leq j \leq d_2 ~  \mbox{and} ~  d_1+j \in S\}$, moreover subset $S^c$ is its complement in $\mathcal{P}$ and  $|S|$ is its cardinality, so that $|\mathcal{P}| =d_1+d_2$. Let us abuse of notations writing $d_1+S^{\bita}=\{d_1+j:  ~ j \in S^{\bita} \}$.
If $S = S^{\balpha} \cup \{ d_1+S^{\bita} \} \subseteq \mathcal{P} $ and $\theta \in \ensuremath{{\mathbb R}}^{d_1+K d_2\times 1}$,  then $\btheta(S)=\left[\balpha\left(S^{\balpha}\right)  ~ \bita\left(S^{\bita}\right)\right] $ denotes reduction of vector $\btheta$ to the subset of group index $S$, as $\balpha\left(S^{\balpha}\right)  \in \ensuremath{{\mathbb R}}^{\left|S^{\balpha}\right| \times 1}$ denotes reduction of vector $\balpha$ to the subset of variable index $S^{\balpha}$ and $\bita\left(S^{\bita}\right)=\left[\left(\bita^{(1)}\left(S^{\bita}\right)\right)^t,  ~ \ldots,  ~\left( \bita^{(K)}\left(S^{\bita}\right)\right)^t\right]^t$ is such that $ \bita^{(k)}\left(S^{\bita}\right) \in \ensuremath{{\mathbb R}}^{\left|S^{\bita}\right| \times 1}$ denotes reduction of vector $\bita^{(k)}$ to the subset of variables index $S^{\bita}$ for all $k=1,\ldots,K$. 
\\

\noindent
\noindent
{\bf Assumptions:}

{\bf (A1)} \ The linear model in Eq. (\ref{eq:general_linear_model})  holds exactly with some \emph{true parameter value} $\btheta_0=\left[\balpha_0^t, \left(\bita^{(1)}_0\right)^t,\ldots, \left(\bita_0^{(K)}\right)^t\right]^t$, $S_0= S^{\balpha}_0 \cup \{ d_1+ S^{\bita}_0 \} $ being the true active set of groups.
\\

\noindent
{\bf (A2)} \ The {\bf compatibility condition} holds for the group index set $S_0= S^{\balpha}_0 \cup \{ d_1+ S^{\bita}_0 \} $ with constant $\phi(S_0) > 0$, if for all $\btheta \in \ensuremath{{\mathbb R}}^{d_1+K d_2 \times 1}$ such that $\| \btheta(S_0^c) \|_{2,1} \leq 3 \| \btheta(S_0) \|_{2,1}$, it holds that
\begin{equation}   \label{eq:comp_cond}
G^{\star}  ~ \left \| \btheta(S_0) \right\|_{2,1}^2   ~ \leq  ~ \left \| \boldsymbol{X} \btheta \right\|_2^2  ~ G^{\star}  ~ |S_0| ~  / ~  nK  ~ \phi(S_0)^2
\end{equation}
\\

Note that Assumption~{\bf (A1)}  means that the true signals $\boldsymbol{c} + \boldsymbol{u}^{(k)} $, for $k=1,\ldots,K$ are exact linear combination of the columns of matrices $\boldsymbol{\Psi}$ and $\boldsymbol{\Phi}$ which simplifies the proof, however this assumption can be relaxed and the following theorem is stated for the best linear approximation of the unknown signals into the span of columns of matrices $\boldsymbol{\Psi}$ and $\boldsymbol{\Phi}$. Moreover, note that in Assumption {\bf  (A2)} $
G^{\star}  ~ |S_0| $ is the average group size times the active number of groups and plays the role of the number of active variables into the compatibility condition. As often observed  the compatibility constant $\phi(S_0)$ is linked to a condition on the smallest eigenvalue of the matrix $\boldsymbol{X}^t \boldsymbol{X} /n$ which turns out to be linked to the product $\boldsymbol{\Phi}^t \boldsymbol{\Psi}$ which in signal processing is the coherence between the two filters. 

We can now prove the following main result:

\begin{theorem} \label{th:oracle_Lasso}

Let $\hat{\btheta}$ be one solution of Eq. (\ref{eq:estimator_grouped}) and let assumptions  {\bf  (A1)} - {\bf  (A2)} hold; then,  
for any $x >0$ and any $\lambda \geq 2 \lambda_0$, with probability at least $1- 2 e^{-x^2/2}-e^{-x}$, 
it holds that
\begin{equation}  \label{eq:oracle}
\frac{1}{nK} \left\| X( \hat{\btheta} - \btheta_0)  \right\|_2^2 +\lambda \sqrt{G^{\star}} \left\| \hat{\btheta} - \btheta_0 \right\|_{2,1} \leq 4   ~  \lambda^2   ~ G^{\star}  ~ |S_0|  / \phi(S_0)^2
\end{equation}
where $\lambda_0=max\left\{\lambda_0^{\balpha}, \lambda_0^{\bita} / \sqrt{K} \right\}$, with
$$
 \lambda_0^{\balpha}= \frac{2 ~ \sigma}{\sqrt{n K}} \sqrt{x^2+ 2 ~ log(d_1)} \quad \mbox{and} \quad \lambda_0^{\bita}= \frac{2 ~ \sigma}{\sqrt{n K}} \left(1 + \sqrt{(4x+4 ~ log(d_2))/K} + (4x+4 ~ log(d_2))/K \right)
$$

\end{theorem}

Proof is given in the Appendix.

\noindent

The theorem proves the so called \emph{ oracle inequality} for the group lasso estimator and it directly gives a bound on the prediction error, indeed if $\lambda$ is chosen as claimed in the theorem, it follows with high probability
$$
\frac{1}{nK} \left\| \boldsymbol{X}( \hat{\btheta} - \btheta_0) \right \|_2^2  \sim \frac{(log(d) ~ \sigma^2 ~ G^{\star} }{nK} ~ |S_0|
$$
with $log(d)=\max\left\{log(d_1), log^2(d_2) / K^2 \right\}$ so that the price for not knowing the true active index groups $S_0$  is of the order $log(d)$.

\section {Simulations and real examples}\label{sec:simul_real}
In order to show the performance of the proposed methodology, a number of experiments were run on synthetic datasets and on a real EEG dataset, the first being an ideal modelization of the second. 

For all results reported in this section, we used the {\tt grpreg} package, that implements efficient algorithms for fitting the regularization path of linear or logistic regression models with different grouped penalties. It includes group selection methods such as group LASSO (referred to as {\tt grlasso} in the following), group MCP, and group SCAD as well as bi-level selection methods such as the group exponential LASSO, the composite MCP, and the group bridge. The smoothing parameter $\lambda$ can be estimated by BIC, AIC, GCV and CV.

We used the group LASSO to solve the penalized regression and the V-fold CV criterion to choose the smoothing parameter $\lambda$.

\subsection{Synthetic data}\label{sec simulations}

In this section we present results obtained using synthetic data representing different sparse scenarios and different noise levels. 
We generated data according to model (\ref{eq:model_channel})
\begin{eqnarray*}
\mathbf{y}^{(k)}=\mathbf{c}+\mathbf{u}^{(k)}+\boldsymbol{\varepsilon}^{(k)}=\boldsymbol{\Psi\alpha}+\boldsymbol{\Phi\beta}^{(k)}+\boldsymbol{\varepsilon}^{(k)} \quad k=1,\dots, K
\end{eqnarray*}  
using three channels ($K=3$) and $n=256$ observations in each channel. 
Matrix $\Psi$ was generated using the following choice 
 $p_{low}=1, \ q_{low}=2, \ s_{low}=1, \ J_{low}=4$ and matrix $\Phi$ was generated using  $p_{high}=8, \ q_{high}=9, \ s_{high}=3, \ J_{high}=10$. These matrices represent  RADWT with  Q-factor almost 1 and 5 respectively, the first frame resembles the dyadic wavelet transform and its mother wavelet has almost one pulse, while the second frame has a mother wavelet with almost 5 pulses, as very well explained in Figure 1 of \cite{Selesnick2011}. We considered three scenarios with different sparsity level:
\begin{enumerate}
\item[] {\it Scenario 1: low sparsity}, corresponding to $\left| S_{\boldsymbol{\alpha}}\right|=24$ and  $\left| S_{\beta}\right|=24$;
\item[] {\it Scenario 2: medium sparsity}, corresponding to $\left| S_{\boldsymbol{\alpha}}\right|=12$ and  $\left| S_{\beta}\right|=12$;
\item[]  {\it Scenario 3: high sparsity}, corresponding to $\left| S_{\boldsymbol{\alpha}}\right|=6$ and  $\left| S_{\beta}\right|=6$;
\end{enumerate}
and for each scenario we used three signal to noise ratios (SNR): 1.5, 3, 6, defined as
$$
\mathrm{SNR}=\frac{\frac{1}{K} \sum_{i=1}^K\mathrm{Var}(\boldsymbol{\Psi\alpha}+\boldsymbol{\Phi\beta}^{(k)})}{\sigma^2_{SNR}}.
$$
Data were generated in each channel, using  $\alpha_{0_j}=1$, $j \in S_0^{\boldsymbol{\alpha}}$, and $\beta_j^{(k)} \sim Uniform(0,M)$, with $M=\left\|\mathbf{c}\right\|_{\infty}/\left\|\boldsymbol{\Phi}(S_0^{\bita})\right\|_{\infty}$, and $\boldsymbol{\varepsilon}^{(k)} \sim N(0,\sigma^2_{\mathrm{SNR}}\boldsymbol{\mathrm{I}})$.
 
In all test cases the proposed procedure, indicated hereafter as {\tt multi-c}, has been compared with the {\tt single-c} procedure, i.e. the procedure where in each channel, the estimator $\mathbf{\hat{f}}^{(k)}=\boldsymbol{\Psi} \hat{\balpha}+\boldsymbol{\Phi} \hat{\bita}^{(k)}$ is obtained independently from the other channels by the following minimization:
$$
\left(\begin{array}{c}
\hat{\balpha} \\
 \hat{\bita}^{(k)}\\
\end{array}\right)  =\underset{ \left(\begin{array}{c}
{\balpha} \\
{\bita}\\
\end{array} \right) \in \ensuremath{{\mathbb R}}^{d_1+d_2 \times 1}}{argmin}  \left\{ \frac{1}{n} \left| \left| \by^{(k)} -[\boldsymbol{\Psi}  ~ \boldsymbol{\Phi}] \left(\begin{array}{c}
{\balpha} \\
{\bita}\\
\end{array} \right)  \right| \right|_2^2 + \lambda  \left(\sum_{i=1}^{d1}|\alpha_i|+ \sum_{i=1}^{d2} |\beta_i| \right)  \right\},
$$
$k=1,\dots,K$.

Performance was evaluated by computing the following indicators:
\begin{itemize}
\item Root Mean Square Error (RMSE) defined as
\begin{equation*}
\mathrm{RMSE}= \sqrt{\frac{1}{n} \sum_{i=1}^{n} \left(\hat{f}^{(k)}(t_i) - f^{(k)}(t_i)\right)^2 },\ k=1,\ldots,K;
\end{equation*}
with $\mathbf{f}^{(k)}=\mathbf{c}+\mathbf{u}^{(k)}$ and $\mathbf{\hat{f}}^{(k)}=\boldsymbol{\Psi} \hat{\balpha}+\boldsymbol{\Phi} \hat{\bita}^{(k)}$ its estimate;
\item Root Mean Square Error for the low resonance component (RMSE$_{low}$) defined as
\begin{equation*}
\mathrm{RMSE}_{low}= \sqrt{\frac{1}{n} \sum_{i=1}^{n} \left(\hat{c}(t_i) - c(t_i)\right)^2 };
\end{equation*}
\item Root Mean Square Error for the high resonance component (RMSE$_{high}$) defined as
\begin{equation*}
\mathrm{RMSE}_{high}= \sqrt{\frac{1}{n} \sum_{i=1}^{n} \left(\hat{u}^{(k)}(t_i) - u^{(k)}(t_i)\right)^2 },\ k=1,\ldots,K;
\end{equation*}
RMSE$_{low}$ and  RMSE$_{high}$ aim at evaluating a component wise accuracy.
\end{itemize}
With the aim of exploring the variable selection properties of the considered procedures, we also computed the following indicators:
\begin{itemize}

\item True positives for the low resonance component (TP$_{low}$)  defined as
\begin{equation*}
\mathrm{TP}_{low}:=\left|\hat S_0^{\boldsymbol{\alpha}}\right|,\ \hat S_0^{\boldsymbol{\alpha}}=\{j: \hat\alpha_{j} \not= 0 \ \mathrm{and} \ \alpha_{0_j} \not= 0  \}.
\end{equation*}
 
\item False negatives for the low resonance component (FN$_{low}$) defined as 
\begin{equation*}
\mathrm{FN}_{low}:= \left|\hat S_0^{\balpha,n}\right|,\ \hat S_0^{\balpha,n}:=\left\{j: \hat \alpha_{j} = 0 \ \mathrm{and} \ \alpha_{0_j}\ne 0\right\}.
\end{equation*}
For the {\tt single-c} procedure TP$_{low}$  and FN$_{low}$ will be dependent on the channels, while for the {\tt multi-c} procedure they will not.
\item True positives for the high resonance component (TP$_{high}$) defined as
\begin{equation*}
\mathrm{TP}_{high}:=\left|\hat S_0^{\bita}\right|=\left|\hat S_0^{(k),\bita}\right|,\ \hat S_0^{(k),\bita}=\left\{j: \hat \beta^{(k)}_{j} \not= 0  \ \mathrm{and} \ \beta^{(k)}_{0_j} \not= 0 \right\},  \quad \forall k=1,\dots,K.
\end{equation*}

\item False negatives for the high resonance component (FN$_{high}$) defined as 
\begin{equation*}
\mathrm{FN}_{high}:= \left|\hat S_0^{\bita,n}\right|= \left|\hat S^{(k),\bita,n}_0\right|,\ \hat S^{(k),\bita,n}_0:=\left\{j: \hat \beta^{(k)}_{j} = 0 \ \mathrm{and} \ \beta^{(k)}_{0_j}\ne 0\right\},  \quad \forall k=1,\dots,K.
\end{equation*}
For the {\tt multi-c} procedure the sets $\hat S_0^{(k),\bita}$ and  $\hat S^{(k),\bita,n}_0$ are all equal, while for the {\tt single-c} procedure the sets depend on the channels.
\end{itemize}
Note that in general the following relationships hold: $\mathrm{TP}=\mathrm{NS}-\mathrm{FP}$ and $\mathrm{TP}+\mathrm{FN}=\mathrm{NS}-\mathrm{FP}+\mathrm{FN}=p_{\mathrm{active}}$, where 
$\mathrm{NS}$ indicates the number of selected variables, $\mathrm{FP}$ indicates the number of false positives and $p_{\mathrm{active}}$ is the true number of active variables.

To be robust with respect to the particular realization in generating synthetic data (and corresponding noise), each experiment was run several times, in particular we set $Nrun=100$ and we evaluated the averaged indicators.

Table \ref{tab ex1} shows the results for RMSE,  RMSE$_{low}$ and  RMSE$_{high}$ for {\it Scenario 1} and SNR= 1.5, 3 and 6 respectively, for all the 3 channels indicated as ch1, ch2, ch3; standard deviation is displayed in parentheses. Table \ref{tab ex1FP_FN} shows the performance indicators TP and FN for the low resonance component and high resonance component.  

Tables \ref{tab ex2}-\ref{tab ex3} contain the results for RMSE,  RMSE$_{low}$ and  RMSE$_{high}$  for {\it Scenario 2} and {\it Scenario 3}, respectively; analogously Tables \ref{tab ex2FP_FN}-\ref{tab ex3FP_FN}  illustrate the performance indicators TP and FN for the same scenarios.

\begin{figure}
\centering
 \includegraphics[width=\textwidth,angle=0]{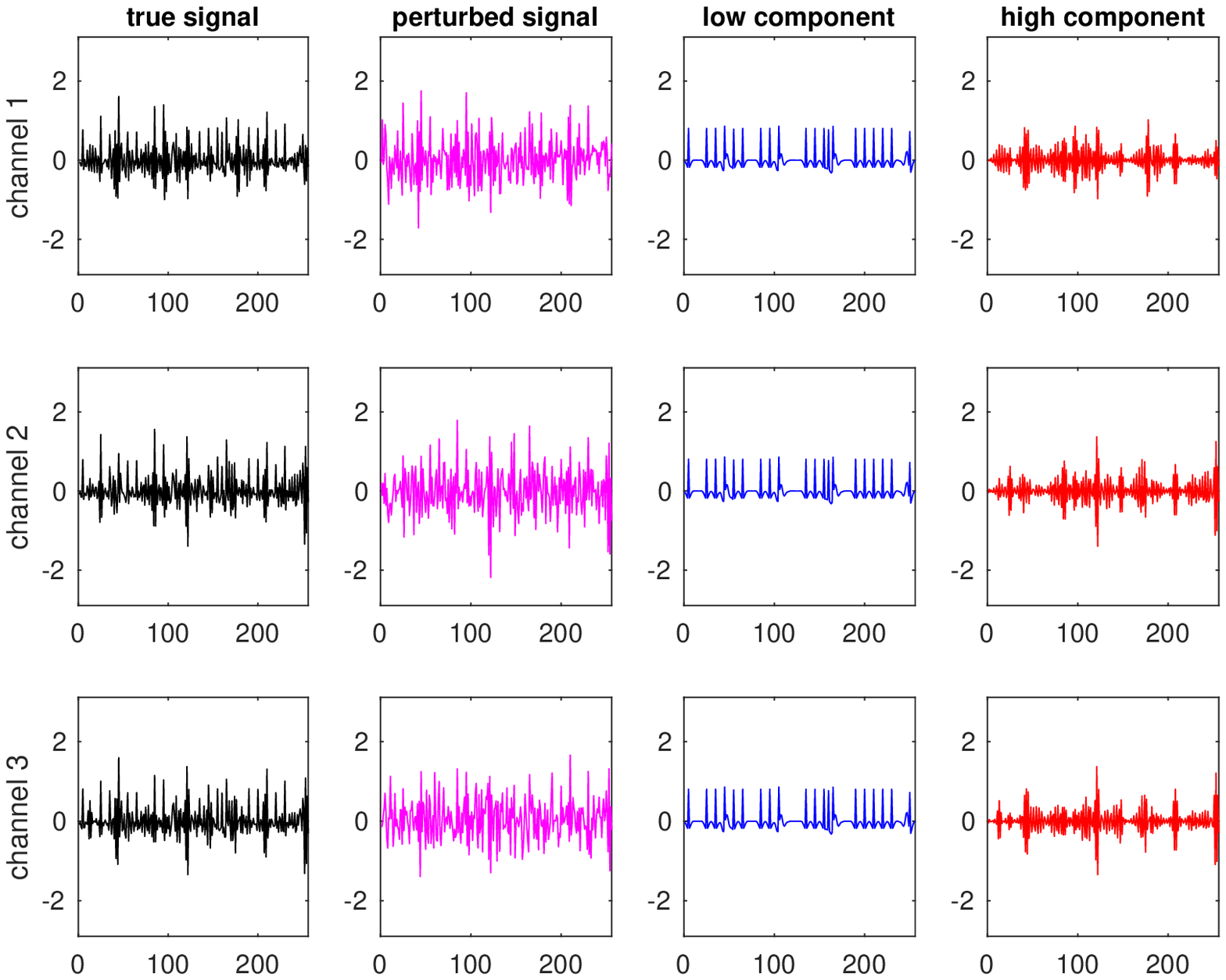}
\caption{True signal (first column),  perturbed signal  for a particular noise realization (second column), true low component (third column), true high component (fourth column) for each channel for {\it Scenario 1} and SNR=1.5.}
\label{true_ex1_snr1.5}
\end{figure}

{\tt Multi-c} procedure always outperforms {\tt single-c} procedure in term of RMSE with a consistently lower standard deviation. This is not surprising because  {\tt multi-c} procedure exploits the joint information among the channels leading to a more precise (mean) and robust (std) estimation error. We also note that, in almost all scenarios and SNRs, {\tt multi-c} outperforms {\tt single-c} reconstructing the two components except for  {\it Scenario 1} where the low and high resonance components share pieces of signals (see Figure \ref{true_ex1_snr1.5}). This is again not surprising, since the two procedures aim to regress $\mathbf{f}=\mathbf{c}+\mathbf{u}$ and not the single components (as in Morphological Component Analysis). Hence,  when the two components low resonance ($\mathbf{c}$) and high resonance ($\mathbf{u}$) are confounding  {\tt single-c} can have some advantage with respect to {\tt multi-c}, remaining the latter more effective in reconstructing the whole signal $f$. The advantage of {\tt multi-c} with respect to {\tt single-c} is more evident looking at the selecting capabilities of the procedure, with a good control of both false positives and false negatives. Of course performance improves when both SNR and sparsity increase. 

For the sake of brevity, we only show the plots of the shape of the unknown signals and the goodness of reconstructions for the two extreme cases, i.e. {\it Scenario 1} with SNR=1.5 and {\it Scenario 3} with SNR=6, see Figures 1-4.

\begin{table}
\caption{ Average values (standard deviation between parentheses) of RMSE, RMSE$_{low}$ and RMSE$_{high}$ based on 100 simulations with different noise realizations. Experiment carried out on Scenario 1 with SNR=1.5, 3 and 6. }
\label{tab ex1}
\resizebox{13cm}{!}{
\begin{tabular}{lcccccc} 
\hline\noalign{\smallskip}
 & \multicolumn{2}{c}{RMSE}&\multicolumn{2}{c}{RMSE$_{low}$}& \multicolumn{2}{c}{RMSE$_{high}$}\\ \cline{2-7}
 & single-c & multi-c & single-c & multi-c & single-c  &multi-c\\ 
\noalign{\smallskip}\hline\noalign{\smallskip}  
SNR=1.5\\
\noalign{\smallskip}\hline
{\tt ch1} & 0.2897 (0.0348) & 0.2216 (0.0191) & 0.2206 (0.0129)  &  0.1728 (0.0127) & 0.2178 (0.0232)  & 0.2284 (0.0154) \\
{\tt ch2} & 0.3004 (0.0355) & 0.2226 (0.0194) & 0.2249 (0.0123) &  0.1728 (0.0127) &  0.2314 (0.0251) & 0.2457 (0.0197)\\
{\tt ch3} & 0.2968 (0.0379) & 0.2130 (0.0187) & 0.2244 (0.0145)  &  0.1728 (0.0127) & 0.2236 (0.0227)  & 0.2337 (0.0194)\\ 
\noalign{\smallskip}\hline
SNR=3\\
\noalign{\smallskip}\hline
{\tt ch1} & 0.2242 (0.0290) & 0.1608 (0.0118) & 0.1882 (0.0170) & 0.1446 (0.0106)  &  0.1715 (0.0156) & 0.1852 (0.0129)   \\
{\tt ch2} & 0.2277 (0.0297) & 0.1628 (0.0113) & 0.1926 (0.0161) & 0.1446 (0.0106) &  0.1842 (0.0175) &  0.2024 (0.0144) \\
{\tt ch3} & 0.2322 (0.0309) & 0.1560 (0.0111) &  0.1924 (0.0165) & 0.1446 (0.0106)  &  0.1815 (0.0175) & 0.1913 (0.0148)  \\ 
\noalign{\smallskip}\hline
SNR=6\\
\noalign{\smallskip}\hline
{\tt ch1} & 0.1611 (0.0234) & 0.1153 (0.0092) & 0.1457 (0.0149)  &  0.1199 (0.0096) & 0.1329 (0.0140)  & 0.1501 (0.0120) \\
{\tt ch2} & 0.1673 (0.0215) & 0.1169 (0.0101) & 0.1554 (0.0129) &  0.1199 (0.0096) &  0.1479 (0.0114) & 0.1615 (0.0138)\\
{\tt ch3} & 0.1613 (0.0233) & 0.1117 (0.0072) & 0.1468 (0.0156)  &  0.1199 (0.0096) & 0.1357 (0.0125)  & 0.1514 (0.0121)\\ 
\noalign{\smallskip}\hline
\end{tabular}
}
\end{table}

\begin{table}
\caption{Fraction of correctly retrieved variables  $\left(\mathrm{TP}_{low}/\left|S_0^{\balpha}\right|\right)$ and incorrectly retrieved variables $\left(\mathrm{FN}_{low}/\left|S_0^{\balpha}\right|\right)$ for the estimated low resonance signal component. Fraction of correctly retrieved variables $\left(\mathrm{TP}_{high}/\left|S_0^{\bita}\right|\right)$ and incorrectly retrieved variables  $\left(\mathrm{FN}_{high}/\left|S_0^{\bita}\right|\right)$  for the estimated high resonance signal component. Values are based on 100 simulations with different noise realizations for Scenario 1 and  SNR=1.5, 3 and 6.}
\label{tab ex1FP_FN}
\resizebox{13cm}{!}{
\begin{tabular}{lcccccccc} 
\hline\noalign{\smallskip}
 & \multicolumn{2}{c}{(TP$_{low}$)/p$_{low}$}&\multicolumn{2}{c}{FN$_{low}$/p$_{low}$} & \multicolumn{2}{c}{(TP$_{high}$)/p$_{high}$}&\multicolumn{2}{c}{FN$_{high}$/p$_{high}$} \\ \cline{2-9}
 & single-c & multi-c & single-c & multi-c & single-c & multi-c& single-c & multi-c\\ 
\noalign{\smallskip}\hline\noalign{\smallskip}  
SNR=1.5\\
\noalign{\smallskip}\hline
{\tt ch1} & 0.4029  & 0.8696  & 0.5971  & 0.1304 & 0.4500 & 0.6508 & 0.5500  & 0.3492  \\
{\tt ch2} & 0.3450  & 0.8696  & 0.6550  & 0.1304  & 0.4129  & 0.6508  & 0.5871 & 0.3492 \\
{\tt ch3} & 0.3629  & 0.8696  & 0.6371  & 0.1304  & 0.4154 & 0.6508  & 0.5846  & 0.3492\\ 
\noalign{\smallskip}\hline
SNR=3\\
\noalign{\smallskip}\hline
{\tt ch1} & 0.6800 & 0.9546  & 0.3200  & 0.0454 & 0.5937 & 0.8613  & 0.4063  & 0.1387   \\
{\tt ch2} & 0.6421  & 0.9546  & 0.3579  & 0.0454 & 0.5767  & 0.8613  & 0.4233  & 0.1387 \\
{\tt ch3} & 0.6662  & 0.9546  & 0.3338  & 0.0454  & 0.5742 & 0.8613  & 0.4258  & 0.1387 \\ 
\noalign{\smallskip}\hline
SNR=6\\
\noalign{\smallskip}\hline
{\tt ch1} & 0.8808 & 0.9912 & 0.1193  & 0.0088 & 0.7137  & 0.9450  & 0.2863  & 0.0550    \\
{\tt ch2} & 0.8487  & 0.9912  & 0.1513  & 0.0088  & 0.6833 & 0.9450  & 0.3167  & 0.0550\\
{\tt ch3} & 0.8775  & 0.9912  & 0.1225  & 0.0088  & 0.7333  & 0.9450 & 0.2667  & 0.0550 \\ 
\noalign{\smallskip}\hline
\end{tabular}
}
\end{table}

\begin{table}
\caption{Average values (standard deviations between parentheses) of RMSE, RMSE$_{low}$ and RMSE$_{high}$ based on 100 simulations with different noise realizations. Experiment carried out on Scenario 2 with SNR=1.5, 3 and 6.}
\label{tab ex2}
\resizebox{13cm}{!}{
\begin{tabular}{lcccccc} 
\hline\noalign{\smallskip}
 & \multicolumn{2}{c}{RMSE}&\multicolumn{2}{c}{RMSE$_{low}$}& \multicolumn{2}{c}{RMSE$_{high}$}\\ \cline{2-7}
 & single-c & multi-c & single-c & multi-c & single-c  &multi-c\\ 
\noalign{\smallskip}\hline\noalign{\smallskip}  
SNR=1.5\\
\noalign{\smallskip}\hline
{\tt ch1} & 0.2151 (0.0187) & 0.1662 (0.0143) & 0.1664 (0.0105)  & 0.1122 (0.0105) & 0.1619 (0.0187)  & 0.1486 (0.0163) \\
{\tt ch2} & 0.2249 (0.0225) & 0.1783 (0.0180) & 0.1646 (0.0116) & 0.1122 (0.0105) &  0.1786 (0.0206) & 0.1660 (0.0188)\\
{\tt ch3} & 0.2175 (0.0197) & 0.1627 (0.0152) & 0.1644 (0.0108)  & 0.1122 (0.0105) & 0.1598 (0.0190)  & 0.1447 (0.0169)\\ 
\noalign{\smallskip}\hline
SNR=3\\
\noalign{\smallskip}\hline
{\tt ch1} & 0.1692 (0.0192) & 0.1209 (0.0114) & 0.1396 (0.0142)  &  0.0826 (0.0078) & 0.1239 (0.0154)  & 0.1099 (0.0130) \\
{\tt ch2} & 0.1748 (0.0184) & 0.1302 (0.0130) & 0.1421 (0.0125) &  0.0826 (0.0078) &  0.1370 (0.0149) & 0.1239 (0.0150)\\
{\tt ch3} & 0.1679 (0.0163) & 0.1154 (0.0099) & 0.1378 (0.0013)  &  0.0826 (0.0078) & 0.1202 (0.0128)  & 0.1038 (0.0120)\\ 
\noalign{\smallskip}\hline
SNR=6\\
\noalign{\smallskip}\hline
{\tt ch1} & 0.1237 (0.0143) & 0.0881 (0.0082) & 0.1059 (0.0126)  &  0.0606 (0.0052) & 0.0944 (0.0098)  & 0.0833 (0.0095) \\
{\tt ch2} & 0.1254 (0.0151) & 0.0941 (0.0089) & 0.1078 (0.0122) &  0.0606 (0.0052) &  0.1047 (0.0102) & 0.0924 (0.0090)\\
{\tt ch3} & 0.1182 (0.0142) & 0.0825 (0.0074) & 0.1004 (0.0138)  &  0.0606 (0.0052) & 0.0891 (0.0093)  & 0.0761 (0.0087)\\ 
\noalign{\smallskip}\hline
\end{tabular}
}
\end{table}

\begin{table}
\caption{Fraction of correctly retrieved variables  $\left(\mathrm{TP}_{low}/\left|S_0^{\balpha}\right|\right)$ and incorrectly retrieved variables $\left(\mathrm{FN}_{low}/\left|S_0^{\balpha}\right|\right)$ for the estimated low resonance signal component. Fraction of correctly retrieved variables $\left(\mathrm{TP}_{high}/\left|S_0^{\bita}\right|\right)$ and incorrectly retrieved variables  $\left(\mathrm{FN}_{high}/\left|S_0^{\bita}\right|\right)$  for the estimated high resonance signal component. Values are based on 100 simulations with different noise realizations for Scenario 2 and  SNR=1.5, 3 and 6.}
\label{tab ex2FP_FN}
\resizebox{13cm}{!}{
\begin{tabular}{lcccccccc} 
\hline\noalign{\smallskip}
 & \multicolumn{2}{c}{(TP$_{low}$)/p$_{low}$}&\multicolumn{2}{c}{FN$_{low}$/p$_{low}$} & \multicolumn{2}{c}{(TP$_{high}$)/p$_{high}$}&\multicolumn{2}{c}{FN$_{high}$/p$_{high}$} \\ \cline{2-9}
 & single-c & multi-c & single-c & multi-c & single-c & multi-c& single-c & multi-c\\ 
\noalign{\smallskip}\hline\noalign{\smallskip}  
SNR=1.5\\
\noalign{\smallskip}\hline
{\tt ch1} & 0.4708  & 0.9508  & 0.5292  & 0.0492  & 0.5708 & 0.7675 & 0.4292  & 0.2325  \\
{\tt ch2} & 0.5017  & 0.9508  & 0.4983  & 0.0492  & 0.6142  & 0.7675  & 0.3858 & 0.2325 \\
{\tt ch3} & 0.4908  & 0.9508  & 0.5092  & 0.0492  & 0.5758 & 0.7675  & 0.4242  & 0.2325 \\ 
\noalign{\smallskip}\hline
SNR=3\\
\noalign{\smallskip}\hline
{\tt ch1} & 0.7867 & 0.9983  & 0.2133  & 0.0017  & 0.6208 & 0.8150 & 0.3792 & 0.1850    \\
{\tt ch2} & 0.7650  & 0.9983 & 0.2350  & 0.0017 & 0.6675 & 0.8150  & 0.3325  & 0.1850 \\
{\tt ch3} & 0.8242  & 0.9983  & 0.1758  & 0.0017  & 0.6608 & 0.8150 & 0.3392  & 0.1850   \\ 
\noalign{\smallskip}\hline
SNR=6\\
\noalign{\smallskip}\hline
{\tt ch1} & 0.9800  & 1 & 0.0200  & 0 & 0.6683 & 0.8508 & 0.3317  & 0.1492  \\
{\tt ch2} & 0.9500  & 1  & 0.0500  & 0  & 0.6808  & 0.8508  & 0.3192 & 0.1492 \\
{\tt ch3} & 0.9725  & 1  & 0.0275  & 0  & 0.7017 & 0.8508  & 0.2983 & 0.1492\\ 
\noalign{\smallskip}\hline
\end{tabular}
}
\end{table}  

\begin{table}
\caption{Average values (standard deviations between parentheses) of RMSE, RMSE$_{low}$ and RMSE$_{high}$ based on 100 simulations with different noise realizations. Experiment carried out on Scenario 3 with SNR=1.5, 3 and 6. }
\label{tab ex3}
\resizebox{13cm}{!}{
\begin{tabular}{lcccccc} 
\hline\noalign{\smallskip}
 & \multicolumn{2}{c}{RMSE}&\multicolumn{2}{c}{RMSE$_{low}$}& \multicolumn{2}{c}{RMSE$_{high}$}\\ \cline{2-7}
 & single-c & multi-c & single-c & multi-c & single-c  &multi-c\\ 
\noalign{\smallskip}\hline\noalign{\smallskip}  
SNR=1.5\\
\noalign{\smallskip}\hline
{\tt ch1} & 0.0437 (0.0104) & 0.0294 (0.0036) & 0.0337 (0.0097)  & 0.0172 (0.0023) & 0.0285 (0.0060)  & 0.0258 (0.0039) \\
{\tt ch2} & 0.0426 (0.0977) & 0.0260 (0.0030) & 0.0347 (0.0093) &  0.0172 (0.0023) &  0.0259 (0.0044) & 0.0218 (0.0034)\\
{\tt ch3} & 0.0459 (0.0103) & 0.0329 (0.0045) & 0.0341 (0.0088)  &  0.0172 (0.0023) & 0.0322 (0.0066)  & 0.0302 (0.0047)\\ 
\noalign{\smallskip}\hline
SNR=3\\
\noalign{\smallskip}\hline
{\tt ch1}   & 0.0298 (0.0053) &  0.0202 (0.0024)  & 0.0228  (0.0045) &  0.0121 (0.0017)  & 0.0205 (0.0039)  & 0.0180 (0.0029)\\
{\tt ch2}  & 0.0283 (0.0048) & 0.0180  (0.0022) & 0.0225  (0.0047)  & 0.0121   (0.0017) & 0.0185  (0.0031) & 0.0151 (0.0028)\\
{\tt ch3}  & 0.0307 (0.0052) & 0.0223 (0.0027) & 0.0226   (0.0043)  & 0.0121  (0.0017) & 0.0223   (0.0041) & 0.0207 (0.0031)\\
\noalign{\smallskip}\hline
SNR=6\\
\noalign{\smallskip}\hline
{\tt ch1} & 0.0201 (0.0033) & 0.0141 (0.0020) & 0.0151 (0.0029) & 0.0084 (0.0013)  &  0.0140 (0.0027) & 0.0126 (0.0023)   \\
{\tt ch2} & 0.0204 (0.0029) & 0.0130 (0.0015) & 0.0156 (0.0026) & 0.0084 (0.0013) &  0.0138 (0.0022) &  0.0113 (0.0017) \\
{\tt ch3} & 0.0217 (0.0032) & 0.0167 (0.0020) & 0.0155 (0.0028) & 0.0084 (0.0013)  &  0.0160 (0.0027) & 0.0151 (0.0022) \\ 
\noalign{\smallskip}\hline
\end{tabular}
}
\end{table}

\begin{table}
\caption{Fraction of correctly retrieved variables  $\left(\mathrm{TP}_{low}/\left|S_0^{\balpha}\right|\right)$ and incorrectly retrieved variables $\left(\mathrm{FN}_{low}/\left|S_0^{\balpha}\right|\right)$ for the estimated low resonance signal component. Fraction of correctly retrieved variables $\left(\mathrm{TP}_{high}/\left|S_0^{\bita}\right|\right)$ and incorrectly retrieved variables  $\left(\mathrm{FN}_{high}/\left|S_0^{\bita}\right|\right)$  for the estimated high resonance signal component. Values are based on 100 simulations with different noise realizations for Scenario 3 and  SNR=1.5, 3 and 6.}
\label{tab ex3FP_FN}
\resizebox{13cm}{!}{
\begin{tabular}{lcccccccc} 
\hline\noalign{\smallskip}
 & \multicolumn{2}{c}{(TP$_{low}$)/p$_{low}$}&\multicolumn{2}{c}{FN$_{low}$/p$_{low}$} & \multicolumn{2}{c}{(TP$_{high}$)/p$_{high}$}&\multicolumn{2}{c}{FN$_{high}$/p$_{high}$} \\ \cline{2-9}
 & single-c & multi-c & single-c & multi-c & single-c & multi-c& single-c & multi-c\\ 
\noalign{\smallskip}\hline\noalign{\smallskip}  
SNR=1.5\\
\noalign{\smallskip}\hline
{\tt ch1}  & 0.9767  & 1  & 0.0233 & 0 & 0.6500 & 0.9833 & 0.3500  & 0.0167  \\
{\tt ch2} & 0.9850  & 1  & 0.0150 & 0  & 0.4300 & 0.9833  & 0.5700 & 0.0167 \\
{\tt ch3} & 0.9800  & 1  & 0.0200 & 0  & 0.8400 & 0.9833  & 0.1600  & 0.0167\\ 
\noalign{\smallskip}\hline
SNR=3\\
\noalign{\smallskip}\hline
{\tt ch1} & 1  & 1  & 0 & 0 & 0.7633 & 1 & 0.2367  & 0  \\
{\tt ch2} & 1  & 1  & 0 & 0  & 0.5867 & 1  & 0.4133 & 0\\
{\tt ch3} & 1  & 1  & 0  & 0  & 0.9933 & 1  & 0.0067  & 0\\ 
\noalign{\smallskip}\hline
SNR=6\\
\noalign{\smallskip}\hline 
{\tt ch1} & 1  & 1  & 0 & 0 & 0.8333 &  1 & 0.1667  & 0  \\
{\tt ch2} & 1  & 1  & 0 & 0 & 0.6633 & 1  & 0.3367 & 0 \\
{\tt ch3} & 1  & 1  & 0 & 0 & 1 & 1  & 0  & 0 \\ 
\noalign{\smallskip}\hline
\end{tabular}
}
\end{table} 

\clearpage

\clearpage

\begin{figure}
\centering
\includegraphics[width=\textwidth,angle=0]{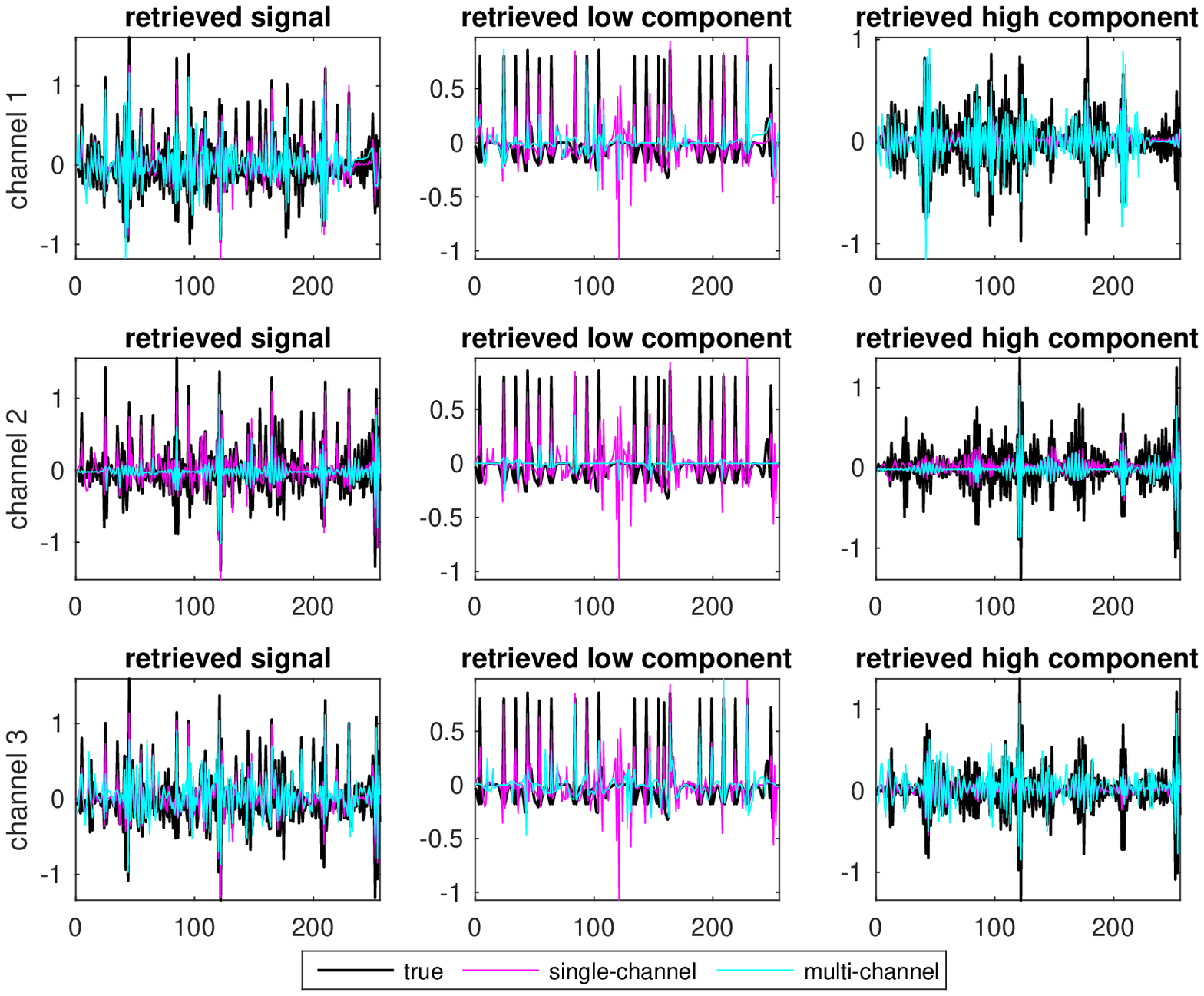}
\caption{Retrieved signal (first column), low component of the retrieved signal (second column), high component of the retrieved signal (third column) for each channel for a particular noise realization, for {\it Scenario 1} and SNR=1.5. Black line refers to the true signal, cyan line refers to single channel retrieval, magenta line refers to multi channel retrieval.}
\label{retr_ex1_snr1.5}
\end{figure}

\clearpage

\begin{figure}
\centering
\includegraphics[width=\textwidth,angle=0]{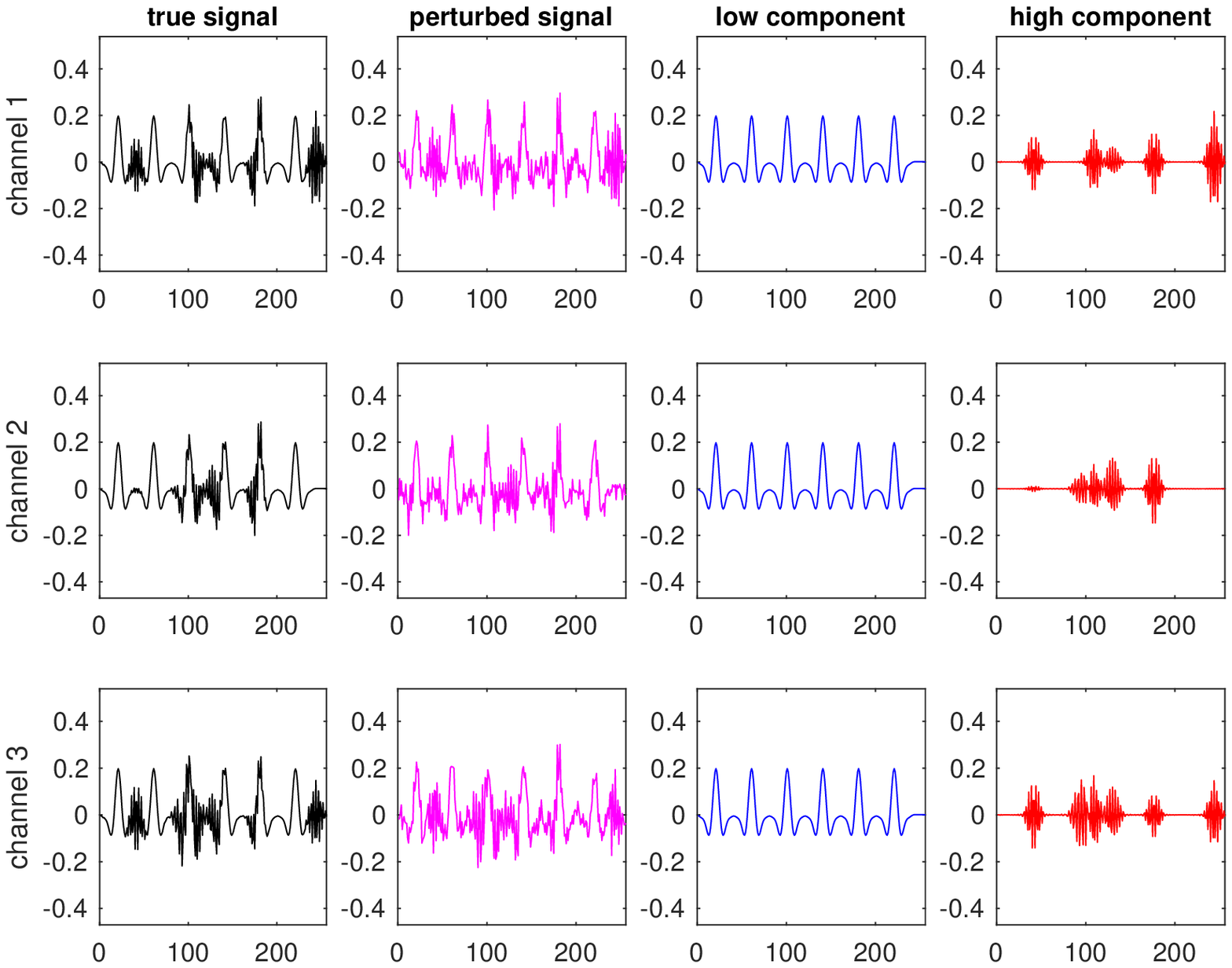}
\caption{True signal (first column),  perturbed signal  for a particular noise realization (second column), true low component (third column), true high component (fourth column) for each channel for {\it Scenario 3} and SNR=6.}
\label{true_ex3_snr6}
\end{figure}

\clearpage

\begin{figure}
\centering
\includegraphics[width=\textwidth,angle=0]{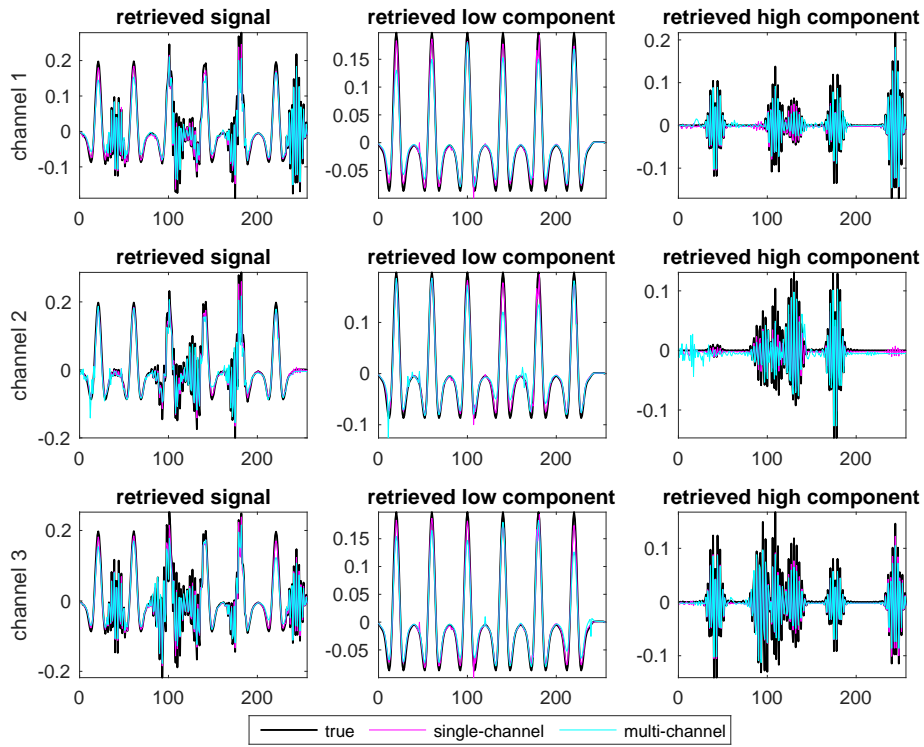}
\caption{Retrieved signal (first column), low component of the retrieved signal (second column), high component of the retrieved signal (third column) for each channel for a particular noise realization,  for {\it Scenario 3} and SNR=6. Black line refers to the true signal, cyan line refers to single channel retrieval, magenta line refers to multi channel retrieval.}
\label{retr_ex3_snr6}
\end{figure}

\clearpage
\subsection{Comparisons and further studies}

For completeness in this section we compare our method with two competitors, namely  BCD and SOMP. These techniques handle multi-task learning problems and their effectiveness
has been shown in diverse survey papers, see \cite{Rakotomamonjy2011} and \citep{Tropp2006_2}. 

The routines mexSOMP and mexL1L2BCD contained in the Matlab SPAMS package (http://spams-devel.gforge.inria.fr/) were used to produce the presented results.
The synthetic data were generated using the same numerical setting of the previous experiment, but we relaxed Hypothesis (H2), setting $\boldsymbol{\beta}^{(3)}=0$. This allowed the data to be different from the correct RADWT model to test the robustness of the method. 

Tables \ref{tab ex1_bias}, \ref{tab ex2_bias} and \ref{tab ex3_bias} show the results of RMSE,  RMSE$_{low}$ and  RMSE$_{high}$  considering SNR= 1.5, 3 and 6, for {\it Scenario 1},  {\it Scenario 2} and  {\it Scenario 3} respectively. The {\tt multi-c} procedure gets a quite significant improvement in terms of RMSE, especially for severe noise condition, mostly due to the good estimation of the low-resonance component. This is not surprising since {\tt multi-c} takes into proper account the equality constraint on the low-component ($\balpha^{(1)}=...=\balpha^{(K)} $). It is also very interesting to note that the {\tt multi-c} procedure outperforms  BCD and SOMP in the retrieval of the high-component of the third channel (which is zero by construction), in fact it gives very low  coefficients $\hat{\beta}^{(3)}$ as properly expected.  

Finally, consistently with the previous analyses, Tables \ref{tab ex1FP_FN_bias}, \ref{tab ex2FP_FN_bias} and \ref{tab ex3FP_FN_bias}  show the performance indicators TP and FN for the low resonance component and high resonance components for {\it Scenario 1},  {\it Scenario 2} and  {\it Scenario 3} respectively. Note that indicators TP and FN are reported only for the first two channels, while for the third channel (which is zero) only the number of falsely non zero retrieved coefficients is reported. It is obvious that, this last index is minimum for the  {\tt single-c} procedure which works on the third channel independently from the other two, however {\tt multi-c} is comparable with SOMP and does a good job with respect to BCD, especially for more severe level of noise.
       
\begin{landscape}
\begin{table}
\caption{ Average values (standard deviations between parentheses) of RMSE, RMSE$_{low}$ and RMSE$_{high}$ based on 100 simulations with different noise realizations. Experiment carried out on Scenario 1 with SNR=1.5, 3 and 6.}
\label{tab ex1_bias}
\resizebox{20cm}{!}{
\begin{tabular}{lcccccccccccc} 
\hline\noalign{\smallskip}
 & \multicolumn{4}{c}{RMSE}&\multicolumn{4}{c}{RMSE$_{low}$}& \multicolumn{4}{c}{RMSE$_{high}$}\\ \cline{2-13}
 & single-c & multi-c & BCD & SOMP & single-c & multi-c & BCD & SOMP & single-c & multi-c & BCD & SOMP \\ 
\noalign{\smallskip}\hline\noalign{\smallskip}  
SNR=1.5\\
\noalign{\smallskip}\hline
{\tt ch1} & 0.2468 (0.0296) & 0.1961 (0.0200) & 0.2035 (0.0097) & 0.2334 (0.0175) & 0.2003 (0.0159) & 0.1172 (0.0092) & 0.1640 (0.0109) & 0.2076 (0.0298) & 0.1790 (0.0185) & 0.1838 (0.0204) & 0.1782 (0.0115) & 0.2147 (0.0281)\\ 
{\tt ch2} & 0.2624 (0.0348) & 0.1953 (0.0169) & 0.1972 (0.0101) & 0.2430 (0.0157) & 0.2099 (0.0172) & 0.1172 (0.0092) & 0.1627 (0.0104) & 0.2190 (0.0283) & 0.1940 (0.0202) & 0.1864 (0.0167) & 0.1634 (0.0131) & 0.2306 (0.0298)\\ 
{\tt ch3} & 0.2280 (0.0258) & 0.1218 (0.0094) & 0.1763 (0.0102) & 0.2268 (0.0171) & 0.2309 (0.0220) & 0.1172 (0.0092) & 0.1574 (0.0112) & 0.1959 (0.0223) & 0.0237 (0.0283) & 0.0455 (0.0135) & 0.1197 (0.0109) & 0.1941 (0.0226)\\ 
\noalign{\smallskip}\hline
SNR=3\\
\noalign{\smallskip}\hline
{\tt ch1} & 0.1893 (0.0239) & 0.1374 (0.0129) & 0.1575 (0.0089) & 0.1577 (0.0109) & 0.1654 (0.0172) & 0.0870 (0.0062) & 0.1258 (0.0085) & 0.1330 (0.0179) & 0.1430 (0.0125) & 0.1330 (0.0133) & 0.1443 (0.0108) & 0.1402 (0.0181)\\ 
{\tt ch2} & 0.1995 (0.0282) & 0.1420 (0.0129) & 0.1435 (0.0088) & 0.1598 (0.0123) & 0.1727 (0.0163) & 0.0870 (0.0062) & 0.1230 (0.0099) & 0.1388 (0.0206) & 0.1561 (0.0152) & 0.1419 (0.0138) & 0.1230 (0.0100) & 0.1474 (0.0198)\\ 
{\tt ch3} & 0.1874 (0.0381) & 0.0925 (0.0065) & 0.1188 (0.0071) & 0.1519 (0.0094) & 0.1926 (0.0360) & 0.0870 (0.0062) & 0.1203 (0.0085) & 0.1260 (0.0170) & 0.0319 (0.0237) & 0.0462 (0.0125) & 0.0709 (0.0080) & 0.1285 (0.0172)\\ 
\noalign{\smallskip}\hline
SNR=6\\
\noalign{\smallskip}\hline
{\tt ch1} & 0.1317 (0.0181) & 0.0968 (0.0072) & 0.1396 (0.0077) & 0.1061 (0.0081) & 0.1213 (0.0141) & 0.0626 (0.0049) & 0.1048 (0.0072) & 0.0874 (0.0132) & 0.1099 (0.0104) & 0.0968 (0.0083) & 0.1255 (0.0086) & 0.0906 (0.0131)\\ 
{\tt ch2} & 0.1361 (0.0211) & 0.1024 (0.0086) & 0.1247 (0.0077) & 0.1088 (0.0083) & 0.1242 (0.0151) & 0.0626 (0.0049) & 0.1020 (0.0067) & 0.0890 (0.0146) & 0.1176 (0.0116) & 0.1059 (0.0096) & 0.1029 (0.0087) & 0.0963 (0.0156)\\ 
{\tt ch3} & 0.1139 (0.0259) & 0.0676 (0.0061) & 0.0943 (0.0062) & 0.1039 (0.0074) & 0.1189 (0.0254) & 0.0626 (0.0049) & 0.1010 (0.0072) & 0.0850 (0.0121) & 0.0401 (0.0198) & 0.0409 (0.0089) & 0.0442 (0.0058) & 0.0860 (0.0107)\\ 
\noalign{\smallskip}\hline
\end{tabular}
}
\end{table}
\end{landscape}

\begin{landscape}
\begin{table}
\caption{Fraction of correctly retrieved variables  $\left(\mathrm{TP}_{low}/\left|S_0^{\balpha}\right|\right)$ and incorrectly retrieved variables $\left(\mathrm{FN}_{low}/\left|S_0^{\balpha}\right|\right)$ for the estimated low resonance signal component. Fraction of correctly retrieved variables $\left(\mathrm{TP}_{high}/\left|S_0^{\bita}\right|\right)$ and incorrectly retrieved variables  $\left(\mathrm{FN}_{high}/\left|S_0^{\bita}\right|\right)$ for channel 1 and 2 and false positives $FP_{high}$ for channel 3 for the estimated high resonance signal component. Values are based on 100 simulations with different noise realizations for Scenario 1 and  SNR=1.5, 3 and 6.}
\label{tab ex1FP_FN_bias}
\resizebox{20cm}{!}{
\begin{tabular}{lcccccccccccccccccccc} 
\hline\noalign{\smallskip}
 & \multicolumn{4}{c}{(TP$_{low}$)/p$_{low}$}&\multicolumn{4}{c}{FN$_{low}$/p$_{low}$} & \multicolumn{4}{c}{(TP$_{high}$)/p$_{high}$} & \multicolumn{4}{c}{FN$_{high}$/p$_{high}$} &\multicolumn{4}{c}{FP$_{high}$}\\ \cline{2-21}
 & single-c & multi-c & BCD & SOMP & single-c & multi-c & BCD & SOMP & single-c & multi-c & BCD & SOMP & single-c & multi-c & BCD & SOMP& single-c & multi-c & BCD & SOMP\\ 
\noalign{\smallskip}\hline\noalign{\smallskip}  
SNR=1.5\\
\noalign{\smallskip}\hline
{\tt ch1} & 0.5713 & 0.9517 & 0.9342 & 0.7792 & 0.4287 & 0.0483 & 0.0658 & 0.2208 & 0.4796 & 0.6667 & 0.7925 & 0.5763 & 0.5204 & 0.3333 & 0.2075 & 0.4238 & - & -& -& -\\ 
{\tt ch2} & 0.5075 & 0.9517 & 0.9342 & 0.7792 & 0.4925 & 0.0483 & 0.0658 & 0.2208 & 0.4638 & 0.6667 & 0.7925 & 0.5763 & 0.5363 & 0.3333 & 0.2075 & 0.4238 & - & -& -& -\\ 
{\tt ch3} & 0.2592 & 0.9517 & 0.9342 & 0.7792 & 0.7408 & 0.0483 & 0.0658 & 0.2208 & - & - & - & - & - & - & - & -& 7.2400 & 42.8300 & 141.4500 & 35.5200\\ 
\noalign{\smallskip}\hline
SNR=3\\
\noalign{\smallskip}\hline
{\tt ch1} & 0.8000 & 0.9938 & 0.9650 & 0.9383 & 0.2000 & 0.0062 & 0.0350 & 0.0617 & 0.5946 & 0.8075 & 0.8492 & 0.7433 & 0.4054 & 0.1925 & 0.1508 & 0.2567 & - & -& -& -\\ 
{\tt ch2} & 0.7896 & 0.9938 & 0.9650 & 0.9383 & 0.2104 & 0.0062 & 0.0350 & 0.0617 & 0.5929 & 0.8075 & 0.8492 & 0.7433 & 0.4071 & 0.1925 & 0.1508 & 0.2567 & - & -& -& -\\ 
{\tt ch3} & 0.6358 & 0.9938 & 0.9650 & 0.9383 & 0.3642 & 0.0062 & 0.0350 & 0.0617 & - & - & - & - & - & - & - & -& 14.6500 & 64.7200 & 101.4200 & 35.7100\\ 
\noalign{\smallskip}\hline
SNR=6\\
\noalign{\smallskip}\hline
{\tt ch1} & 0.9238 & 0.9996 & 0.9888 & 0.9888 & 0.0762 & 0.0004 & 0.0113 & 0.0113 & 0.7438 & 0.9062 & 0.8896 & 0.8554 & 0.2563 & 0.0938 & 0.1104 & 0.1446 & - & -& -& -\\ 
{\tt ch2} & 0.9450 & 0.9996 & 0.9888 & 0.9888 & 0.0550 & 0.0004 & 0.0113 & 0.0113 & 0.7200 & 0.9062 & 0.8896 & 0.8554 & 0.2800 & 0.0938 & 0.1104 & 0.1446 & - & -& -& -\\ 
{\tt ch3} & 0.9446 & 0.9996 & 0.9888 & 0.9888 & 0.0554 & 0.0004 & 0.0113 & 0.0113 & - & - & - & - & - & - & - & -& 34.0200 & 80.3600 & 62.5500 & 35.7900\\ 
\noalign{\smallskip}\hline
\end{tabular}
}
\end{table}   
\end{landscape}

\begin{landscape}
\begin{table}
\caption{Average values (standard deviations between parentheses) of RMSE, RMSE$_{low}$ and RMSE$_{high}$ based on 100 simulations with different noise realizations. Experiment carried out on Scenario 2 with SNR=1.5, 3 and 6.}
\label{tab ex2_bias}
\resizebox{20cm}{!}{
\begin{tabular}{lcccccccccccc} 
\hline\noalign{\smallskip}
 & \multicolumn{4}{c}{RMSE}&\multicolumn{4}{c}{RMSE$_{low}$}& \multicolumn{4}{c}{RMSE$_{high}$}\\ \cline{2-13}
 & single-c & multi-c & BCD & SOMP & single-c & multi-c & BCD & SOMP & single-c & multi-c & BCD & SOMP \\ 
\noalign{\smallskip}\hline\noalign{\smallskip}  
SNR=1.5\\
\noalign{\smallskip}\hline
{\tt ch1} & 0.1848 (0.0183) & 0.1560 (0.0167) & 0.1555 (0.0108) & 0.1520 (0.0159) & 0.1478 (0.0127) & 0.0834 (0.0082) & 0.1180 (0.0110) & 0.1093 (0.0210) & 0.1497 (0.0156) & 0.1521 (0.0179) & 0.1301 (0.0137) & 0.1275 (0.0202)\\ 
{\tt ch2} & 0.1795 (0.0211) & 0.1360 (0.0126) & 0.1494 (0.0097) & 0.1486 (0.0141) & 0.1455 (0.0129) & 0.0834 (0.0082) & 0.1122 (0.0109) & 0.1102 (0.0201) & 0.1318 (0.0160) & 0.1237 (0.0135) & 0.1204 (0.0105) & 0.1212 (0.0176)\\ 
{\tt ch3} & 0.1673 (0.0188) & 0.0898 (0.0087) & 0.1353 (0.0106) & 0.1439 (0.0165) & 0.1674 (0.0187) & 0.0834 (0.0082) & 0.1111 (0.0111) & 0.1077 (0.0215) & 0.0158 (0.0194) & 0.0414 (0.0115) & 0.0878 (0.0096) & 0.1114 (0.0184)\\ 
\noalign{\smallskip}\hline
SNR=3\\
\noalign{\smallskip}\hline
{\tt ch1} & 0.1374 (0.0153) & 0.1094 (0.0105) & 0.1113 (0.0084) & 0.1027 (0.0117) & 0.1165 (0.0122) & 0.0603 (0.0061) & 0.0851 (0.0077) & 0.0723 (0.0151) & 0.1103 (0.0111) & 0.1069 (0.0112) & 0.0964 (0.0108) & 0.0808 (0.0149)\\ 
{\tt ch2} & 0.1307 (0.0155) & 0.1001 (0.0101) & 0.1079 (0.0091) & 0.1020 (0.0102) & 0.1106 (0.0131) & 0.0603 (0.0061) & 0.0819 (0.0079) & 0.0720 (0.0130) & 0.0995 (0.0106) & 0.0938 (0.0105) & 0.0905 (0.0091) & 0.0825 (0.0127)\\ 
{\tt ch3} & 0.1214 (0.0227) & 0.0659 (0.0060) & 0.0844 (0.0087) & 0.0977 (0.0091) & 0.1222 (0.0231) & 0.0603 (0.0061) & 0.0792 (0.0099) & 0.0710 (0.0120) & 0.0204 (0.0165) & 0.0347 (0.0070) & 0.0483 (0.0064) & 0.0745 (0.0117)\\ 
\noalign{\smallskip}\hline
SNR=6\\
\noalign{\smallskip}\hline
{\tt ch1} & 0.0945 (0.0103) & 0.0790 (0.0077) & 0.0951 (0.0077) & 0.0692 (0.0068) & 0.0822 (0.0094) & 0.0442 (0.0046) & 0.0703 (0.0067) & 0.0502 (0.0083) & 0.0794 (0.0086) & 0.0793 (0.0081) & 0.0802 (0.0083) & 0.0536 (0.0088)\\ 
{\tt ch2} & 0.0912 (0.0090) & 0.0731 (0.0075) & 0.0934 (0.0073) & 0.0702 (0.0076) & 0.0791 (0.0097) & 0.0442 (0.0046) & 0.0670 (0.0071) & 0.0500 (0.0090) & 0.0755 (0.0073) & 0.0713 (0.0083) & 0.0775 (0.0073) & 0.0555 (0.0081)\\ 
{\tt ch3} & 0.0804 (0.0152) & 0.0483 (0.0047) & 0.0645 (0.0069) & 0.0688 (0.0077) & 0.0780 (0.0159) & 0.0442 (0.0046) & 0.0637 (0.0076) & 0.0493 (0.0096) & 0.0261 (0.0183) & 0.0276 (0.0061) & 0.0278 (0.0053) & 0.0532 (0.0092)\\ 
\noalign{\smallskip}\hline
\end{tabular}
}
\end{table}
\end{landscape}

\begin{landscape}
\begin{table}
\caption{Fraction of correctly retrieved variables  $\left(\mathrm{TP}_{low}/\left|S_0^{\balpha}\right|\right)$ and incorrectly retrieved variables $\left(\mathrm{FN}_{low}/\left|S_0^{\balpha}\right|\right)$ for the estimated low resonance signal component. Fraction of correctly retrieved variables $\left(\mathrm{TP}_{high}/\left|S_0^{\bita}\right|\right)$ and incorrectly retrieved variables  $\left(\mathrm{FN}_{high}/\left|S_0^{\bita}\right|\right)$ for channel 1 and 2 and false positives $FP_{high}$ for channel 3 for the estimated high resonance signal component. Values are based on 100 simulations with different noise realizations for Scenario 2 and  SNR=1.5, 3 and 6.}
\label{tab ex2FP_FN_bias}
\resizebox{20cm}{!}{
\begin{tabular}{lcccccccccccccccccccc} 
\hline\noalign{\smallskip}
 & \multicolumn{4}{c}{(TP$_{low}$)/p$_{low}$}&\multicolumn{4}{c}{FN$_{low}$/p$_{low}$} & \multicolumn{4}{c}{(TP$_{high}$)/p$_{high}$} & \multicolumn{4}{c}{FN$_{high}$/p$_{high}$} &\multicolumn{4}{c}{FP$_{high}$}\\ \cline{2-21}
 & single-c & multi-c & BCD & SOMP & single-c & multi-c & BCD & SOMP & single-c & multi-c & BCD & SOMP & single-c & multi-c & BCD & SOMP& single-c & multi-c & BCD & SOMP\\ 
\noalign{\smallskip}\hline\noalign{\smallskip}  
SNR=1.5\\
\noalign{\smallskip}\hline
{\tt ch1} & 0.6983 & 0.9992 & 0.9925 & 0.9550 & 0.3017 & 0.0008 & 0.0075 & 0.0450 & 0.7825 & 0.8575 & 0.8308 & 0.7692 & 0.2175 & 0.1425 & 0.1692 & 0.2308 & - & -& -& -\\ 
{\tt ch2} & 0.7242 & 0.9992 & 0.9925 & 0.9550 & 0.2758 & 0.0008 & 0.0075 & 0.0450 & 0.5642 & 0.8575 & 0.8308 & 0.7692 & 0.4358 & 0.1425 & 0.1692 & 0.2308 & - & -& -& -\\ 
{\tt ch3} & 0.4233 & 0.9992 & 0.9925 & 0.9550 & 0.5767 & 0.0008 & 0.0075 & 0.0450 & - & - & - & - & - & - & - & -& 5.0300 & 35.9400 & 120.7500 & 18.8000\\ 
\noalign{\smallskip}\hline
SNR=3\\
\noalign{\smallskip}\hline
{\tt ch1} & 0.9333 & 1.0000 & 1.0000 & 0.9983 & 0.0667 & 0.0000 & 0.0000 & 0.0017 & 0.8158 & 0.9050 & 0.8350 & 0.8758 & 0.1842 & 0.0950 & 0.1650 & 0.1242 & - & -& -& -\\ 
{\tt ch2} & 0.9608 & 1.0000 & 1.0000 & 0.9983 & 0.0392 & 0.0000 & 0.0000 & 0.0017 & 0.6567 & 0.9050 & 0.8350 & 0.8758 & 0.3433 & 0.0950 & 0.1650 & 0.1242 & - & -& -& -\\ 
{\tt ch3} & 0.9133 & 1.0000 & 1.0000 & 0.9983 & 0.0867 & 0.0000 & 0.0000 & 0.0017 & - & - & - & - & - & - & - & -& 11.6100 & 38.2500 & 73.4700 & 18.8700\\ 
\noalign{\smallskip}\hline
SNR=6\\
\noalign{\smallskip}\hline
{\tt ch1} & 0.9983 & 1.0000 & 1.0000 & 1.0000 & 0.0017 & 0.0000 & 0.0000 & 0.0000 & 0.8292 & 0.9267 & 0.8350 & 0.9325 & 0.1708 & 0.0733 & 0.1650 & 0.0675 & - & -& -& -\\ 
{\tt ch2} & 0.9983 & 1.0000 & 1.0000 & 1.0000 & 0.0017 & 0.0000 & 0.0000 & 0.0000 & 0.6908 & 0.9267 & 0.8350 & 0.9325 & 0.3092 & 0.0733 & 0.1650 & 0.0675 & - & -& -& -\\ 
{\tt ch3} & 0.9983 & 1.0000 & 1.0000 & 1.0000 & 0.0017 & 0.0000 & 0.0000 & 0.0000 & - & - & - & - & - & - & - & -& 27.5800 & 42.7800 & 35.3700 & 18.9900\\ 
\noalign{\smallskip}\hline
\end{tabular}
}
\end{table}   
\end{landscape}

\begin{landscape}
\begin{table}
\caption{Average values (standard deviations between parentheses) of RMSE, RMSE$_{low}$ and RMSE$_{high}$ based on 100 simulations with different noise realizations. Experiment carried out on Scenario 3 with SNR=1.5, 3 and 6.}
\label{tab ex3_bias}
\resizebox{20cm}{!}{
\begin{tabular}{lcccccccccccc} 
\hline\noalign{\smallskip}
 & \multicolumn{4}{c}{RMSE}&\multicolumn{4}{c}{RMSE$_{low}$}& \multicolumn{4}{c}{RMSE$_{high}$}\\ \cline{2-13}
 & single-c & multi-c & BCD & SOMP & single-c & multi-c & BCD & SOMP & single-c & multi-c & BCD & SOMP \\ 
\noalign{\smallskip}\hline\noalign{\smallskip}  
SNR=1.5\\
\noalign{\smallskip}\hline
{\tt ch1} & 0.0382 (0.0077) & 0.0296 (0.0038) & 0.0291 (0.0029) & 0.0294 (0.0036) & 0.0299 (0.0071) & 0.0154 (0.0023) & 0.0195 (0.0032) & 0.0197 (0.0047) & 0.0248 (0.0049) & 0.0262 (0.0040) & 0.0225 (0.0032) & 0.0225 (0.0045)\\ 
{\tt ch2} & 0.0395 (0.0077) & 0.0259 (0.0029) & 0.0259 (0.0028) & 0.0305 (0.0039) & 0.0319 (0.0071) & 0.0154 (0.0023) & 0.0194 (0.0031) & 0.0195 (0.0052) & 0.0244 (0.0042) & 0.0217 (0.0029) & 0.0180 (0.0028) & 0.0238 (0.0050)\\ 
{\tt ch3} & 0.0324 (0.0072) & 0.0162 (0.0023) & 0.0204 (0.0028) & 0.0281 (0.0035) & 0.0321 (0.0076) & 0.0154 (0.0023) & 0.0192 (0.0031) & 0.0195 (0.0055) & 0.0050 (0.0048) & 0.0038 (0.0021) & 0.0081 (0.0019) & 0.0198 (0.0050)\\ 
\noalign{\smallskip}\hline
SNR=3\\
\noalign{\smallskip}\hline
{\tt ch1} & 0.0276 (0.0051) & 0.0206 (0.0029) & 0.0260 (0.0023) & 0.0195 (0.0031) & 0.0213 (0.0045) & 0.0111 (0.0013) & 0.0164 (0.0026) & 0.0129 (0.0039) & 0.0183 (0.0033) & 0.0183 (0.0031) & 0.0203 (0.0024) & 0.0147 (0.0035)\\ 
{\tt ch2} & 0.0283 (0.0046) & 0.0193 (0.0020) & 0.0235 (0.0019) & 0.0207 (0.0026) & 0.0222 (0.0042) & 0.0111 (0.0013) & 0.0167 (0.0023) & 0.0132 (0.0038) & 0.0183 (0.0030) & 0.0167 (0.0021) & 0.0166 (0.0019) & 0.0163 (0.0028)\\ 
{\tt ch3} & 0.0236 (0.0045) & 0.0120 (0.0014) & 0.0166 (0.0026) & 0.0195 (0.0027) & 0.0235 (0.0047) & 0.0111 (0.0013) & 0.0164 (0.0027) & 0.0126 (0.0038) & 0.0030 (0.0028) & 0.0041 (0.0015) & 0.0036 (0.0013) & 0.0147 (0.0037)\\ 
\noalign{\smallskip}\hline
SNR=6\\
\noalign{\smallskip}\hline
{\tt ch1} & 0.0194 (0.0031) & 0.0150 (0.0019) & 0.0256 (0.0018) & 0.0136 (0.0017) & 0.0145 (0.0029) & 0.0076 (0.0011) & 0.0160 (0.0016) & 0.0090 (0.0023) & 0.0135 (0.0024) & 0.0136 (0.0022) & 0.0200 (0.0019) & 0.0103 (0.0023)\\ 
{\tt ch2} & 0.0200 (0.0032) & 0.0146 (0.0014) & 0.0228 (0.0017) & 0.0140 (0.0021) & 0.0147 (0.0030) & 0.0076 (0.0011) & 0.0159 (0.0018) & 0.0090 (0.0026) & 0.0139 (0.0022) & 0.0131 (0.0017) & 0.0163 (0.0015) & 0.0108 (0.0025)\\ 
{\tt ch3} & 0.0170 (0.0035) & 0.0084 (0.0011) & 0.0160 (0.0019) & 0.0136 (0.0018) & 0.0170 (0.0036) & 0.0076 (0.0011) & 0.0159 (0.0019) & 0.0092 (0.0025) & 0.0023 (0.0023) & 0.0033 (0.0013) & 0.0021 (0.0008) & 0.0101 (0.0025)\\ 
\noalign{\smallskip}\hline
\end{tabular}
}
\end{table}
\end{landscape}

\begin{landscape}
\begin{table}
\caption{Fraction of correctly retrieved variables  $\left(\mathrm{TP}_{low}/\left|S_0^{\balpha}\right|\right)$ and incorrectly retrieved variables $\left(\mathrm{FN}_{low}/\left|S_0^{\balpha}\right|\right)$ for the estimated low resonance signal component. Fraction of correctly retrieved variables $\left(\mathrm{TP}_{high}/\left|S_0^{\bita}\right|\right)$ and incorrectly retrieved variables  $\left(\mathrm{FN}_{high}/\left|S_0^{\bita}\right|\right)$ for channel 1 and 2 and false positives $FP_{high}$ for channel 3 for the estimated high resonance signal component. Values are based on 100 simulations with different noise realizations for Scenario 3 and  SNR=1.5, 3 and 6.}
\label{tab ex3FP_FN_bias}
\resizebox{20cm}{!}{
\begin{tabular}{lcccccccccccccccccccc} 
\hline\noalign{\smallskip}
 & \multicolumn{4}{c}{(TP$_{low}$)/p$_{low}$}&\multicolumn{4}{c}{FN$_{low}$/p$_{low}$} & \multicolumn{4}{c}{(TP$_{high}$)/p$_{high}$} & \multicolumn{4}{c}{FN$_{high}$/p$_{high}$} &\multicolumn{4}{c}{FP$_{high}$}\\ \cline{2-21}
 & single-c & multi-c & BCD & SOMP & single-c & multi-c & BCD & SOMP & single-c & multi-c & BCD & SOMP & single-c & multi-c & BCD & SOMP& single-c & multi-c & BCD & SOMP\\ 
\noalign{\smallskip}\hline\noalign{\smallskip}  
SNR=1.5\\
\noalign{\smallskip}\hline
{\tt ch1} & 0.9900 & 1.0000 & 1.0000 & 1.0000 & 0.0100 & 0.0000 & 0.0000 & 0.0000 & 0.5467 & 0.6033 & 0.8483 & 0.6467 & 0.4533 & 0.3967 & 0.1517 & 0.3533 & - & -& -& -\\ 
{\tt ch2} & 1.0000 & 1.0000 & 1.0000 & 1.0000 & 0.0000 & 0.0000 & 0.0000 & 0.0000 & 0.4983 & 0.6033 & 0.8483 & 0.6467 & 0.5017 & 0.3967 & 0.1517 & 0.3533 & - & -& -& -\\ 
{\tt ch3} & 0.9967 & 1.0000 & 1.0000 & 1.0000 & 0.0033 & 0.0000 & 0.0000 & 0.0000 & - & - & - & - & - & - & - & -& 7.3300 & 8.3600 & 35.7400 & 8.4200\\ 
\noalign{\smallskip}\hline
SNR=3\\
\noalign{\smallskip}\hline
{\tt ch1} & 1.0000 & 1.0000 & 1.0000 & 1.0000 & 0.0000 & 0.0000 & 0.0000 & 0.0000 & 0.6517 & 0.7950 & 0.8717 & 0.7933 & 0.3483 & 0.2050 & 0.1283 & 0.2067 & - & -& -& -\\ 
{\tt ch2} & 1.0000 & 1.0000 & 1.0000 & 1.0000 & 0.0000 & 0.0000 & 0.0000 & 0.0000 & 0.6633 & 0.7950 & 0.8717 & 0.7933 & 0.3367 & 0.2050 & 0.1283 & 0.2067 & - & -& -& -\\ 
{\tt ch3} & 1.0000 & 1.0000 & 1.0000 & 1.0000 & 0.0000 & 0.0000 & 0.0000 & 0.0000 & - & - & - & - & - & - & - & -& 4.7100 & 13.9000 & 9.2600 & 9.0000\\ 
\noalign{\smallskip}\hline
SNR=6\\
\noalign{\smallskip}\hline
{\tt ch1} & 1.0000 & 1.0000 & 1.0000 & 1.0000 & 0.0000 & 0.0000 & 0.0000 & 0.0000 & 0.7433 & 0.9200 & 0.8500 & 0.9133 & 0.2567 & 0.0800 & 0.1500 & 0.0867 & - & -& -& -\\ 
{\tt ch2} & 1.0000 & 1.0000 & 1.0000 & 1.0000 & 0.0000 & 0.0000 & 0.0000 & 0.0000 & 0.7833 & 0.9200 & 0.8500 & 0.9133 & 0.2167 & 0.0800 & 0.1500 & 0.0867 & - & -& -& -\\ 
{\tt ch3} & 1.0000 & 1.0000 & 1.0000 & 1.0000 & 0.0000 & 0.0000 & 0.0000 & 0.0000 & - & - & - & - & - & - & - & -& 5.9600 & 13.6600 & 5.6400 & 9.2700\\ 
\noalign{\smallskip}\hline
\end{tabular}
}
\end{table}   
\end{landscape}

\clearpage
\subsection{Real data}

To illustrate our procedure in a real case, we considered the problem of separating the transient and the oscillatory component in human sleep EEG data. This problem is actually a very hot topic in neuroscience, because several studies have pointed out the benefit of separating the transients and oscillations before spindle detection, see \cite{Coppieters2016} and \cite{Parekh2015}. There exist already several methods for separating transients and oscillations in EEG data, but here we refer to \cite{Lajnef2015} where the joint detection of sleep spindles and K-complex events are obtained using a Morphological Component Analysis (MCA) and two different RADWT with respectively high and low Q-factor, as supposed in this paper. On the other hand, although the American Academy of Sleep Medicne (AASM) manual recommends using more the one channel for scoring sleep and associated events, actually only few available methods advocate the use of multichannel EEG (\cite{Barros2000}, \cite{Parekh2017}), then our procedure can be considered a possible alternative in this respect.
 
In particular in this section we show results obtained by applying our proposed multichannel procedure to one publicly sleep EEG database, the DREAMS Sleep Spindles Database available at www.tcts.fpms.ac.be/$\sim$devuyst/Databases/DatabaseSpindles/. This database has been produced by the University of MONS - TCTS Laboratory (Stéphanie Devuyst, Thierry Dutoit) and the Université Libre de Bruxelles - CHU de Charleroi Sleep Laboratory (Myriam Kerkhofs).

These data were acquired in a sleep laboratory of a Belgium hospital using a digital 32-channel polygraph (BrainnetTM System of MEDATEC, Brussels, Belgium). They consist of height polysomnographic recordings coming from patients with different pathologies (dysomnia, restless legs syndrome, insomnia, apnoea/hypopnoea syndrome). Two EOG channels (P8-A1, P18-A1), three EEG channels (CZ-A1 or C3-A1, FP1-A1 and O1-A1) and one submental EMG channel were recorded. The standard European Data Format (EDF) was used for storing. The sampling frequency was 200Hz, 100Hz or 50Hz. A segment of 30 minutes of the central EEG channel was extracted from each whole-night recording for spindles scoring, giving origin to 8 excerpts of 30 minutes. No effort was made to select good spindle epochs or noise free epochs, in order to reflect reality as much as possible. These excerpts were given independently to two experts for sleep spindles scoring.

In particular we focus on excerpt2 sampled at 200Hz extracted from 00:00:00 to 00:30:00 with annotated EEG channels CZ-A1, FP1-A1 and O1-A1, belonging to a 40-years man, i.e. 3 signals, one for channel, formed by 360000 time points.

\begin{figure}
\centering
\includegraphics[width=\textwidth,angle=0]{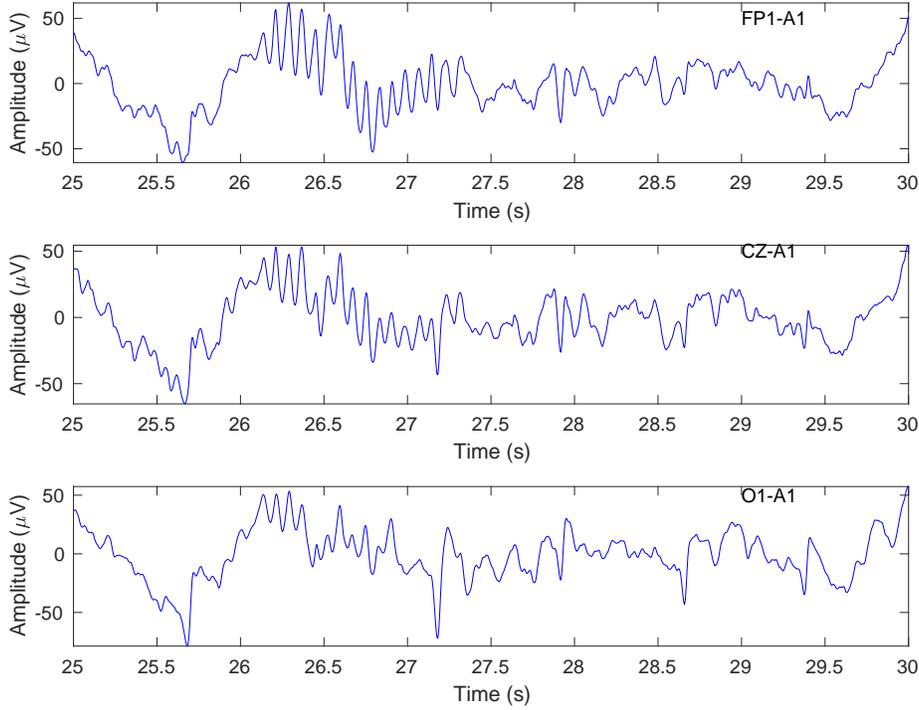}
\caption{EEG segment 6 corresponding to time interval 25-30 sec for the 3 EEG channels, FP1-A1 , CZ-A1 and O1-A1.}
\label{signal_6}
\end{figure}

\begin{figure}
\centering
\includegraphics[width=\textwidth,angle=0]{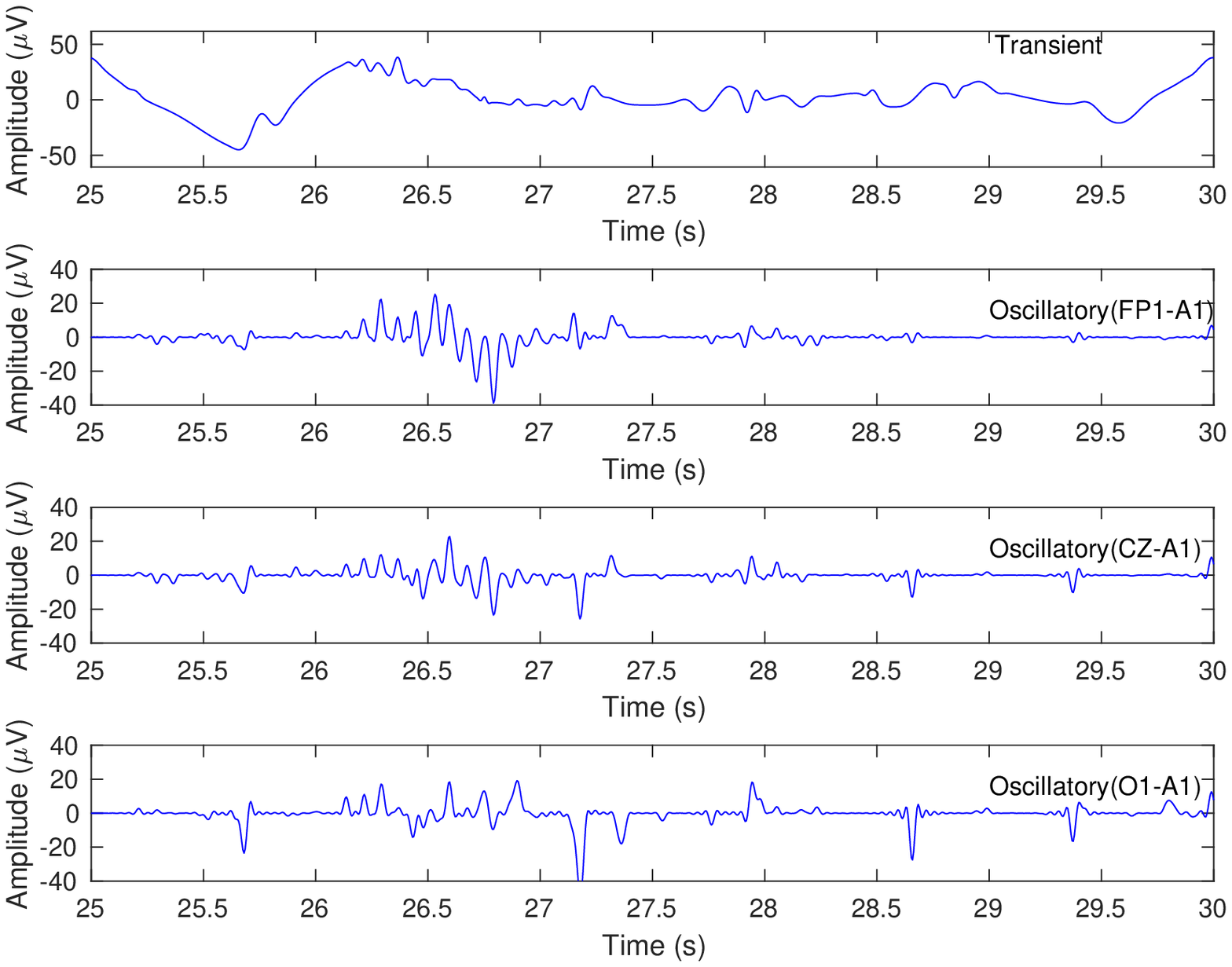}
\caption{Retrieved low-transient and high-oscillatory components of segment 6 corresponding to time interval 25-30 sec for the 3 EEG channels, FP1-A1 , CZ-A1 and O1-A1.}
\label{low_high_6}
\end{figure}

\begin{figure}
\centering
\includegraphics[width=\textwidth,angle=0]{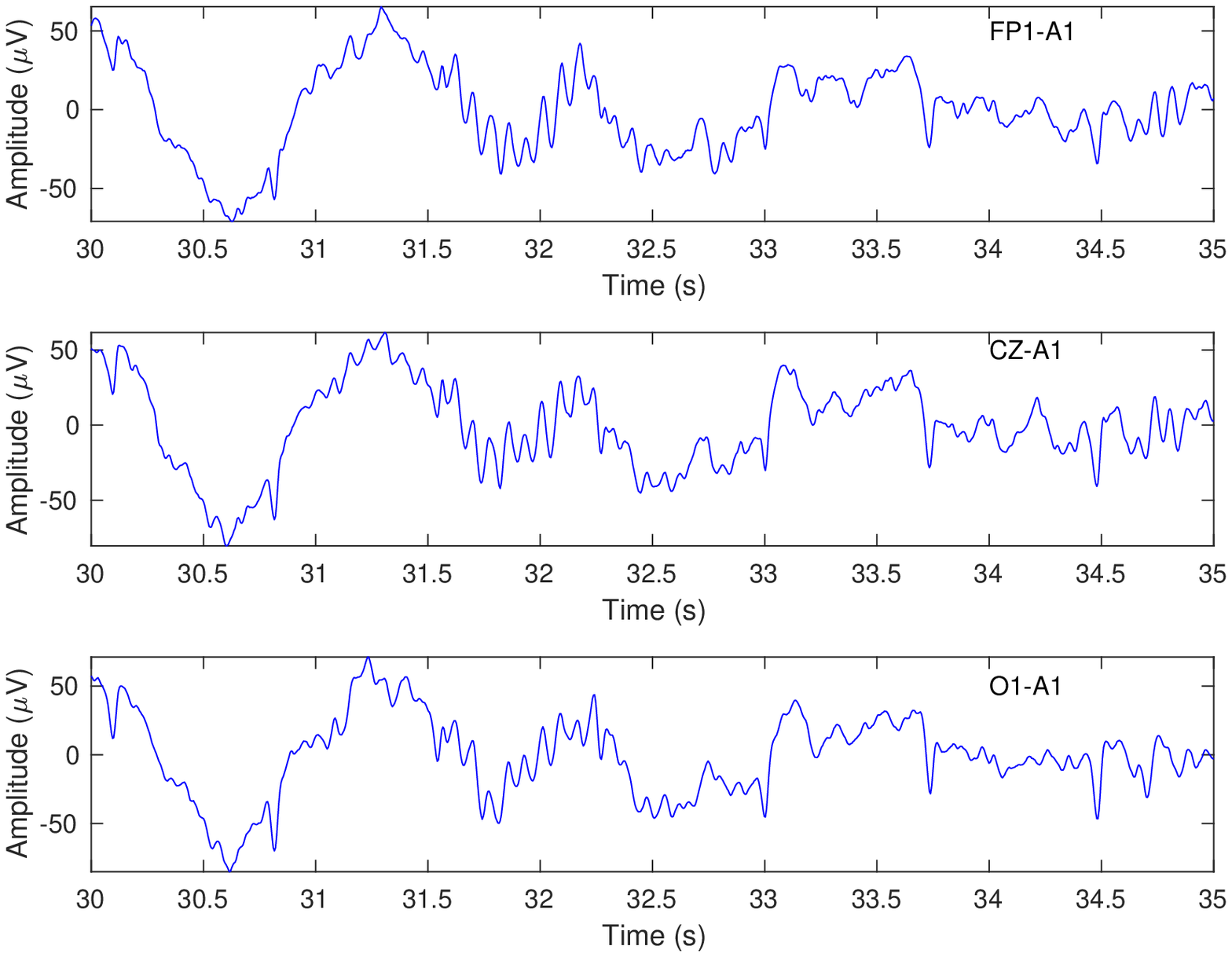}
\caption{EEG segment 7 corresponding to time interval 30-35 sec for the 3 EEG channels, FP1-A1 , CZ-A1 and O1-A1.}
\label{signal_7}
\end{figure}

We segmented each signal in 360 segments of length 1000 time points, corresponding to 5 seconds, and we concentrate only on the 200 segments corresponding to sleep phase 2. In particular we focused on two consecutive segments: 25-30 sec and 30-35 sec, see Figures \ref{signal_6}-\ref{signal_7} respectively. In both the segments the two experts annotated visually spindles events at same times. Indeed, in the first segment, the first expert annotated a spindle event at 26.09 sec of length 1.28 sec and the second expert annotated the event at 26.12 sec with length 1 sec; in the second segment, the first expert annotated a spindle event at 31.5 sec of length 0.74 sec and the second expert annotated the event at 31.515 sec with length 1 sec.

Following \cite{Lajnef2015}, we suppose the oscillatory part to be well described by an RADWT with Q-factor=5 (which roughly corresponds to the choice $p=8, q=9, s=3, J=10$) and the transient part to be well represented by an RADWT with Q-factor=1 (which roughly corresponds to the choice $p=1, q=2, s=1, J=4$). Moreover we suppose that hypothesis (H1) is true, since we are considering sleep data where the epochs containing electrode artifacts due to lead and other body movements are not analyzed, hence we expect the 3 channels share the same underground/transient activity; we also suppose that hypothesis (H2) is true, since the spindles events, which represent the major and also the most interesting contribution to the oscillating part, simultaneously activate in the 3 channels, as widely discussed in \cite{Parekh2017}.


\begin{figure}
\centering
\includegraphics[width=\textwidth,angle=0]{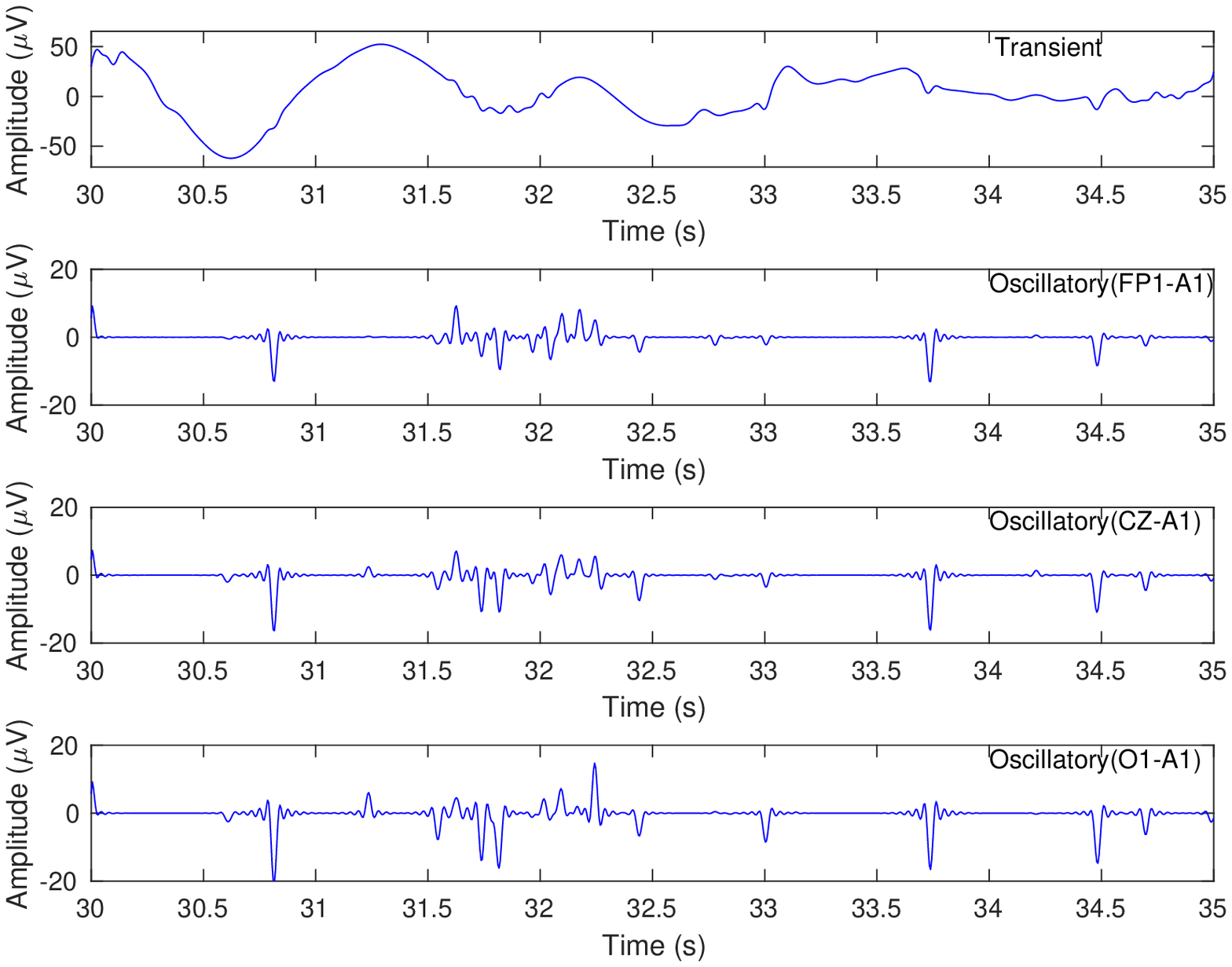}
\caption{Retrieved low-transient and high-oscillatory components  of segment 7 corresponding to time interval 25-30 sec for the 3 EEG channels, FP1-A1 , CZ-A1 and O1-A1.}
\label{low_high_7}
\end{figure}
Figures \ref{low_high_6}-\ref{low_high_7}  show the retrieval of the transient and oscillatory components for the two considered segments, 25-30 sec and 30-35 sec respectively.
From the figures we can see how the transient part is really faithful to the underlying trend of the three channels, it keeps some oscillations that do not persist in time; moreover we can appreciate the 3 oscillatory components, in which similar but not equal oscillations resonate in the same time intervals. These phenomena correspond to the spindle events that most likely occur contemporaneously on the three EEG channels with similar characteristics being not exactly the same. Of course this procedure must be considered as a preprocessing step for an automatic spindles detection, which in this case appear very clear around sec. 26  in the first excerpt (visually inspecting Figure \ref{low_high_6}) and around sec. 32 in the second excerpt (visually inspecting Figure \ref{low_high_7}).
The analyzed segment 25-30 sec correspond to the segment analyzed in paper \cite{Parekh2017}, see Figure 5, and it can be seen that the position of spindles coincides.


\section{Conclusions}
 \label{sec:conclus}
In this paper we presented a method for nonparametric regression analysis of multichannel signals under a structural hypothesis on the underlying signals covering some specific real life situations. The method leverages on a complete filter bank (RADWT) that defines a frame in $L_2(R)$ which guarantees a perfect reconstruction property and a tunable Q-factor. In our work we used two frames, one with low Q-factor and one with high Q-factor, able to represent sparsity of signals with low and high resonance respectively. The structural hypothesis on the underlying signals explicitly states that in each channel the signal is a sum of two contributions, one (the low resonance signal) is common to all channels, while the other (the high resonance signal) is channel-specific but retains the same spectral properties in each channel, i.e. the positions of non-zero RADWT coefficients. We showed the connections with the SSA problem, stressing the difference between our proposal and the existing literature.

Firstly, we applied the method on a set of synthetic data satisfying the mathematical hypotheses, showing its ability in retrieving the signal in each channel, as expected from its asymptotic properties. We also compared its performance with other two techniques proposed in the literature, namely SOMP and BCD, considering a second synthetic dataset from a non correct RADWT generative model to test the robustness. Moreover, we displayed its skill in reconstructing the individual components and in controlling the sparsity of the model too. Finally, the proposed technique was tested on  human sleep EEG data, confirming some results already studied in the literature.  

Future research is devoted to the improvement of the algorithm in pursuing component specific results.

 \section*{Appendix}
 \label{sec:append}
Before proving Theorem \ref{th:oracle_Lasso}, let us present some  preliminary results.
 
\vspace{4mm}
 
\noindent

For each $j=1,\ldots,d_1$, define the random variables 
\begin{equation}\label{eq:uj}
u_j= \frac{\bepsilon^t \boldsymbol{X}^{(j)}}{\sqrt{n K}} \quad  \quad  \quad  \quad  \quad \mbox{with} \quad \boldsymbol{X}^{(j)}=\left[\underbrace{\left(\boldsymbol{\Psi}^{(j)}\right)^t, \cdots, \left(\boldsymbol{\Psi}^{(j)}\right)^t }_{K  ~ times} \right]^t,
\end{equation}
where $\boldsymbol{X}^{(j)}$ is the $j-$th column of matrix $\boldsymbol{X}$ and $\boldsymbol{\Psi}^{(j)}$  the $j-$th column of matrix $\boldsymbol{\Psi}$.

\vspace{4mm}
 
\noindent

{\bf Proposition 1:} \ For the random variables $u_j$ it holds for any $x >0$
\begin{equation}
P \left( \underset{1 \leq j \leq d_1}{max} 2 |u_j|  < \sqrt{n K} \lambda_0^{\alpha} \right)  \geq 1-2 e^{-x^2/2}
\end{equation} 
where 
$$
\lambda_0^{\alpha}= \frac{2 ~ \sigma}{\sqrt{n K}} \sqrt{x^2+ 2 ~ log(d_1)}
$$

Proof: since $u_j=\frac{1}{\sqrt{nK}} \sum_{k=1}^K \sum_{i=1}^n \epsilon_i^{(k)} \Psi^{(j)}_i \sim \mathcal{N}(0,\sigma^2)$  we can apply lemma 6.2 of \cite{geer_book} and result is proved.

\vspace{4mm}
 
\noindent

For each $j=1,\ldots,d_2$, define the random variables
\begin{equation} \label{eq:vj}
v_j= \frac{\left\|\bepsilon^t \tilde{\boldsymbol{X}}^{(j)} \right\|_2}{\sqrt{n K}}   \quad  \quad  \quad  \quad  \quad \mbox{with} \quad 
\tilde{\boldsymbol{X}}^{(j)}=\left[ \begin{array}{cccc}
\boldsymbol{\Phi}^{(j)} & \boldsymbol{0} & \cdots & \boldsymbol{0} \\
\boldsymbol{0}& \boldsymbol{\Phi}^{(j)}  & \cdots & \boldsymbol{0} \\
\cdots & & & \cdots \\
\boldsymbol{0} & \boldsymbol{0} & \cdots & \boldsymbol{\Phi}^{(j)}  \\
\end{array}   ~
 \right] 
\end{equation}

being  a matrix of dimension $n K \times K$ with $\boldsymbol{\Phi}^{(j)}$ the $j-$th column of matrix $\boldsymbol{\Phi}$.

\vspace{4mm}
 
\noindent

{\bf Proposition 2:} \ For the random variables $v_j$ it holds for any $x >0$
\begin{equation}
P \left( \underset{1 \leq j \leq d_12}{max} 2 |v_j|  < \sqrt{n K} \lambda_0^{\bita} \right)  \geq 1-e^{-x}
\end{equation} 
where 
$$
\lambda_0^{\bita}= \frac{2 ~ \sigma}{\sqrt{n K}} \left(1 + \sqrt{(4x+4 ~ log(d_2))/K} + (4x+4 ~ log(d_2))/K \right)
$$

Proof: by definition we have that
$$
v_j=\frac{1}{\sqrt{K}} \left| \left| \left( \frac{\sum_{i=1}^{n}  \epsilon^{(1)}_i \Phi^{(j)}_i}{\sqrt{n}},   \cdots,  \frac{\sum_{i=1}^{n}  \epsilon^{(K)}_i \Phi^{(j)}_i}{\sqrt{n}} \right) \right|\right|_2 = \frac{\sigma}{\sqrt{K}}  \left( \sum_{k=1}^K \left( \frac{\sum_{i=1}^{n}  \epsilon^{(k)}_i \Phi^{(j)}_i}{\sigma \sqrt{n}} \right)^2 \right)^{1/2}.
$$
Since  $\frac{\sum_{i=1}^{n}  \epsilon^{(k)}_i \Phi^{(j)}_i}{\sigma \sqrt{n}}$ are $K$ independent normal standard variables, we have that  $ K v^2_j / \sigma^2 \sim \chi^2(K)$. Finally,  applying  lemma 8.1 of \cite{geer_book} result is proved.

\vspace{4mm}
 
\noindent

{\bf Proposition 3:} \ For all $\btheta \in \ensuremath{{\mathbb R}}^{d_1+K d_2 \times 1}$ and for any $x>0$ it holds 
$$
P \left( \frac{2  ~  \bepsilon^t ~  \boldsymbol{X}   ~ \btheta }{n K} \leq \lambda_0 \sqrt{G^{\star}} \| \btheta \|_{2,1} \right) \geq 1-2 e^{-x^2/2} - e^{-x}
$$

with $\bepsilon$ the concatenation of noise vectors given in Eq. (\ref{eq:general_linear_model}) and $\lambda_0 =max\{\lambda_0^{\balpha}, \lambda_0^{\bita} / \sqrt{K} \}$.

Proof: by definitions of $\btheta$  we can write

$$
 \frac{2  ~  \bepsilon^t ~  \boldsymbol{X}  ~ \btheta }{n K}=\frac{2  ~  \bepsilon^t ~  \boldsymbol{X}   }{n K} ~  \left[\begin{array}{c}
\balpha \\
\bita^{(1)} \\
\bita^{(2)} \\
\vdots \\
\bita^{(K)} \\
\end{array}   \right] =\frac{1}{\sqrt{nK}} \left( \sum_{j=1}^{d_1} 2 \frac{\bepsilon^t \boldsymbol{X}^{(j)}}{\sqrt{n K}}  \alpha_j +\sum_{j=1}^{d_2} \left(  \frac{ \bepsilon^t \tilde{\boldsymbol{X}}^{(j)} }{\sqrt{n K}} \right) \bita_j^{(\cdot)}  \right) 
$$
where $\bita_j^{(\cdot)}=\left[\beta_j^{(1)},...,\beta_j^{(K)}\right]^t$, while $\boldsymbol{X}^{(j)}$ and $ \tilde{\boldsymbol{X}}^{(j)}$ are given in (\ref{eq:uj}) and (\ref{eq:vj}). Using Proposition 1 and 2 and the fact that $u v \leq |u| |v| $, $\forall \; u,v \in \mathbb{R} $ and  $<\boldsymbol{u},\boldsymbol{v}> \leq \|\boldsymbol{u}\|_2 \|\boldsymbol{v}\|_2$, $\forall \; \boldsymbol{u}, \boldsymbol{v} \in \mathbb{R}^K$,  with probability at least $1-2 e^{-x^2/2} - e^{-x}$ it follows
\begin{eqnarray*}
 \frac{2  ~  \bepsilon^t ~  \boldsymbol{X}   ~ \btheta }{n K} & \leq & \frac{1}{\sqrt{nK}} \left( \sum_{j=1}^{d_1} 2 \frac{\left|\bepsilon^t \boldsymbol{X}^{(j)}\right|}{\sqrt{n K}}  |\alpha_j| +2\sum_{j=1}^{d_2}  \frac{\left\| \bepsilon^t \tilde{\boldsymbol{X}}^{(j)} \right\|_2}{\sqrt{n K}}  \left\|\bita_j^{(\cdot)}\right\|_2  \right)  \\
& \leq &  \frac{1}{\sqrt{nK}} \left( \sum_{j=1}^{d_1} 2 |u_j| |\alpha_j| +\sum_{j=1}^{d_2} 2|v_j| \left\| \bita_j^{(\cdot)}\right\|_2  \right) \\
& \leq &  \frac{1}{\sqrt{nK}} \left( \underset{1 \leq j \leq d_1}{max} 2 |u_j|   ~ \|\balpha\|_1 +   \underset{1 \leq j \leq d_1}{max} 2 |v_j|     ~ \sum_{j=1}^{d_2} \left\|\bita_j^{(\cdot)}\right\|_2  \right) \\
& \leq &   \left( \lambda_0^{\balpha}   ~ \|\balpha\|_1 +  \frac{\lambda_0^{\bita}}{\sqrt{K}} \sqrt{K}     ~ \sum_{j=1}^{d_2} \left\|\bita_j^{(\cdot)}\right\|_2  \right) \\
& \leq &  \sqrt{G^{\star}} \lambda_0  \left( \frac{1}{\sqrt{G^{\star}}}\|\balpha\|_1 + \sqrt{\frac{K}{G^{\star}}}   ~ \sum_{j=1}^{d_2} \left\|\bita_j^{(\cdot)}\right\|_2  \right) =  \sqrt{G^{\star}} \lambda_0   ~ \|  \btheta \|_{2,1},
\end{eqnarray*}
where $\lambda_0=max\{\lambda_0^{\balpha}, \lambda_0^{\bita}  / \sqrt{K} \}$.

\vspace{8mm}
 
\noindent

{\bf Proof of Theorem \ref{th:oracle_Lasso}:}

By definition of $\hat{\btheta}$ and $\btheta_0$  it holds 
$$
 \frac{1}{nK} \left\|  \by -\boldsymbol{X} \hat{\btheta} \right\| _2^2 + \lambda \sqrt{G^{\star}} \left\| \hat{\btheta}\right\| _{2,1} \leq  \frac{1}{nK} \left\|  \by -\boldsymbol{X} \btheta_0 \right\| _2^2 + \lambda \sqrt{G^{\star}} \left\| \btheta_0\right\| _{2,1} ,
$$
then, by using $\by=\boldsymbol{X} \btheta_0 + \bepsilon$,  it also holds 
$$
\frac{1}{nK} \left\| \boldsymbol{X} ( \hat{\btheta}-\btheta_0)\right\| _2^2 + \lambda \sqrt{G^{\star}} \left\| \hat{\btheta}\right\| _{2,1} \leq  \frac{2 \bepsilon^t \boldsymbol{X} ( \hat{\btheta}-\btheta_0)}{n K}  + \lambda \sqrt{G^{\star}} \left\| \btheta_0 \right\| _{2,1}.
$$
Chose any $x$, then with probability at least $1-2 e^{-x^2/2} - e^{-x}$, by Proposition 3, it holds
$$
\frac{1}{nK} \left\| \boldsymbol{X} ( \hat{\btheta}-\btheta_0) \right\| _2^2 + \lambda \sqrt{G^{\star}} \left\| \hat{\btheta}\right\| _{2,1} \leq \sqrt{G^{\star}} \lambda_0 \| ( \hat{\btheta}-\btheta_0)  \|_{2,1} + \lambda \sqrt{G^{\star}} \left\| \btheta_0\right\| _{2,1}.
$$
Chose $ \lambda > 2 \lambda_0$, and observe that, whatever $S_0 \subseteq \mathcal{P}$, one has $\left\| \btheta \right\| _{2,1} = \left\| \btheta(S_0)\right\| _{2,1} +\left\|  \btheta(S^c_0) \right\| _{2,1}$ for any $\btheta$ and in particular $\left\|  \btheta_0 \right\| _{2,1}= \left\|  \btheta_0(S_0)\right\| _{2,1}$, then it holds 
$$
\frac{2}{nK} \left\| \boldsymbol{X} ( \hat{\btheta}-\btheta_0) \right\| _2^2 + 2 \lambda \sqrt{G^{\star}} \left\| \hat{\btheta}(S_0^c)-\btheta_0(S_0^c)\right\| _{2,1} \leq \sqrt{G^{\star}}  \lambda \left\| \hat{\btheta}-\btheta_0 \right\| _{2,1} +2 \lambda \sqrt{G^{\star}} \left(\left\| \btheta_0(S_0)\right\| _{2,1} -\left\| \hat{\btheta}(S_0)\right\| _{2,1} \right).
$$
By using the triangle inequality  for the $l_2/l_1-$norm, $\left| ~   \| v \|_{2,1} -  \| u\|_{2,1} ~  \right| \leq \| u-v \|_{2,1}$ and rewriting $\left\| \hat{\btheta}-\btheta_0 \right\|_{2,1}=\left\| \hat{\btheta}(S_0)-\btheta_0(S_0) \right\|_{2,1}+\left\| \hat{\btheta}(S_0^c)-\btheta_0 (S_0^c)\right\|_{2,1}$, it holds 

\begin{equation}  \label{eq:intermidiate1}
\frac{2}{nK} \left\| \boldsymbol{X} ( \hat{\btheta}-\btheta_0) \right\| _2^2 + \lambda \sqrt{G^{\star}} \left\| \hat{\btheta}(S_0^c)-\btheta_0(S_0^c)\right\| _{2,1} \leq  3 \lambda \sqrt{G^{\star}} \left\| \hat{\btheta}(S_0)-\btheta_0(S_0)\right\| _{2,1}.
\end{equation}

Now from Eq. (\ref{eq:intermidiate1}) we obtain two consequences. The first is that $\left\| \hat{\btheta}(S_0^c)-\btheta_0(S_0^c)\right\| _{2,1} \leq  3 \left\| \hat{\btheta}(S_0)-\btheta_0(S_0)\right\| _{2,1}$, hence for assumption {\bf(A2)}, it holds

\begin{equation}  \label{eq:intermidiate2}
G^{\star}\left\| \hat{\btheta}(S_0)-\btheta_0(S_0)\right\| ^2_{2,1} \leq  \frac{\left\| \boldsymbol{X} ( \hat{\btheta}-\btheta_0)\right\| _2^2  ~  G^{\star}   |S_0|  }{nK  \phi(S_0)^2}.
\end{equation}
The second is obtained adding $\lambda \sqrt{G^{\star}} \left\| \hat{\btheta}(S_0)-\btheta_0(S_0)\right\| _{2,1}$ on both sides of Eq. (\ref{eq:intermidiate1}), hence

\begin{equation}  \label{eq:intermidiate3}
\frac{2}{nK} \left\| \boldsymbol{X} ( \hat{\btheta}-\btheta_0) \right\| _2^2 + \lambda \sqrt{G^{\star}} \left\| \hat{\btheta}-\btheta_0\right\| _{2,1} \leq  4 \lambda \sqrt{G^{\star}} \left\| \hat{\btheta}(S_0)-\btheta_0(S_0)\right\| _{2,1}.
\end{equation}
Now, substitute Eq. (\ref{eq:intermidiate2}) into Eq. (\ref{eq:intermidiate3}) and obtain

$$
\frac{2}{nK} \left\| \boldsymbol{X} ( \hat{\btheta}-\btheta_0) \right\| _2^2 + \lambda \sqrt{G^{\star}} \left\| \hat{\btheta}-\btheta_0\right\| _{2,1} \leq  4 \lambda \frac{\left\| \boldsymbol{X} ( \hat{\btheta}-\btheta_0) \right\| _2  ~  \sqrt{G^{\star}  ~  |S_0| }    }{\sqrt{nK}  ~   \phi(S_0)}.
$$
Finally, using the inequality $4 u v \leq u^2 + 4 v^2$, we obtain
$$
\frac{2}{nK}\left\| \boldsymbol{X} ( \hat{\btheta}-\btheta_0) \right\| _2^2 + \lambda \sqrt{G^{\star}} \left\| \hat{\btheta}-\btheta_0\right\| _{2,1} \leq    \frac{\left\| \boldsymbol{X} ( \hat{\btheta}-\btheta_0) \right\| _2^2 }{nK} + 4 \frac{\lambda^2 ~  G^{\star}  ~  |S_0| }{ \phi(S_0)^2},
$$
which gives Eq. (\ref{eq:oracle}).

\section*{References}

\bibliography{mybibfile}

\section*{Acknowledgements}
Daniela De Canditiis was partially supported by grant INdAM-GNCS Project 2018.

\end{document}